\documentclass{pas}

\usepackage{graphicx}	
\usepackage{algorithm,algcompatible}
\usepackage{braket}
\usepackage{xcolor}
\usepackage{float, makecell, array}
\usepackage{xspace, bm}
\usepackage{pdflscape}
\usepackage{longtable, makecell}
\usepackage{subfigure}
\usepackage{multirow, multicol}
\usepackage{fontawesome} 
\usepackage[dvipsnames]{xcolor}
\usepackage{scalerel}
\usepackage{enumerate}
\usepackage{chngcntr}

\usepackage{lipsum}
\usepackage{makecell}

\newcommand{\gold}{\textcolor[HTML]{AB8F25}{gold}\xspace}

\newcommand{\hbeta}{H$\mathrm{\beta}$\xspace}
\newcommand{\mgii}{Mg\textsc{ii}\xspace}
\newcommand{\civ}{C\textsc{iv}\xspace}
\newcommand{\feii}{Fe\textsc{ii}\xspace}

\newcommand{\hbetamath}{\text{H}\beta}

\newcommand{\ev}{\mathcal{Z}}

\newcommand{\beq}{\begin{equation}}
\newcommand{\eeq}{\end{equation}}
\newcommand{\bea}{\begin{eqnarray}}
\newcommand{\eea}{\end{eqnarray}}
\newcommand{\multilinecomment}[1]{}

\newcommand{\qm}[1]{``#1''}
\newcommand{\sqm}[1]{`#1'}
\newcommand{\hescan}{Laplace Quadrature\xspace}
\newcommand{\sviscan}{SVI Quadrature\xspace}

\newcommand{\abs}[1]{\stretchleftright{|}{#1}{|}}
\newcommand{\BF}[1]{\mathrm{BF}_\mathrm{#1}}

\newcommand{\pone}{$16$\textsuperscript{th}}
\newcommand{\ptwo}{$84$\textsuperscript{th}}
\newcommand{\redraw}{\mathrm{Redraw}}

\newcommand{\ESSo}{\text{ESS}_1}
\newcommand{\ESSt}{\text{ESS}_2}
\newcommand{\fidp}{\emptyset}

\newcommand{\dayu}{\xspace\mathrm{ \ d}}
\newcommand{\angstrom}{\xspace\mathrm{ \ \AA}}

\newcommand{\red}{\textcolor{red}}

\newcommand{\des}{DES\xspace}
\newcommand{\decam}{DECam\xspace}
\newcommand{\ozdes}{OzDES\xspace}
\newcommand{\sdss}{SDSS\xspace}
\newcommand{\lsst}{LSST\xspace}
\newcommand{\desi}{DESI\xspace}

\newcommand{\tides}{TiDES\xspace}

\newcommand{\javelin}{\texttt{JAVELIN}\xspace}
\newcommand{\emcee}{\texttt{emcee}\xspace}

\newcommand{\mica}{\texttt{MICA}\xspace}
\newcommand{\litmus}{\texttt{LITMUS}\xspace}
\newcommand{\python}{\texttt{python}\xspace}

\newcommand{\numpyro}{\texttt{numpyro}\xspace}
\newcommand{\jax}{\texttt{jax}\xspace}

\newcommand{\jaxns}{\texttt{JAXNS}\xspace}

\usepackage{natbib}
\usepackage{orcidlink}

\DeclareUnicodeCharacter{2009}{ }
\DeclareUnicodeCharacter{1E3A}{L} 
\DeclareUnicodeCharacter{223C}{$\sim$}

\newcommand{\tmax}{2500 \dayu}
\newcommand{\paperone}{HM1\xspace}
\newcommand{\papertwo}{HM2\xspace}
\newcommand{\paperthree}{HM3\xspace}

\definecolor{color1}{HTML}{1f77b4}
\definecolor{color2}{HTML}{ff7f0e}
\definecolor{color3}{HTML}{2ca02c}
\definecolor{color4}{HTML}{d62728}
\hypersetup{
    colorlinks,
    linkcolor={color1},
    citecolor={color1},
    urlcolor={color1}
}

\begin{document}
\lefttitle{Publications of the Astronomical Society of Australia}
\righttitle{McDougall et al.}

\jnlPage{1}{4}
\jnlDoiYr{2026}
\doival{10.1017/pasa.xxxx.xx}

\articletitt{Research Paper}

\title{The Radius-Luminosity Relationship for Active Galactic Nuclei with Hierarchical Bayesian Reverberation Mapping}

\author{
\href{https://orcid.org/0009-0008-5846-1543}{\gn{Hugh} \sn{McDougall}}$^{1}$,
\href{https://orcid.org/0000-0002-4213-8783}{\gn{Tamara M.} \sn{Davis}}$^{1}$, 
\href{https://orcid.org/0000-0003-2595-9114}{\gn{Benjamin~J.~S. } \sn{Pope}}$^{2,3}$,  
\href{https://orcid.org/0000-0002-4279-4182}{\gn{Paul} \sn{Martini}}$^{4,5}$, 
\href{https://orcid.org/0000-0003-0644-9282}{\gn{Zhefu} \sn{Yu}}$^{6}$, 
\href{https://orcid.org/0000-0003-1731-0497}{\gn{Chris} \sn{Lidman}}$^{7,8}$, 
\href{https://orcid.org/0000-0003-3081-9319}{\gn{Geraint~F.} \sn{Lewis}}$^{9}$,
\href{https://orcid.org/0000-0001-5525-6255}{\gn{Christian} \sn{Adamcewicz}}$^{10}$,
\href{https://orcid.org/0009-0008-3940-5992}{\gn{Rahma} \sn{Alfarsy}}$^{11}$
}

\affil{
$^{1}$School of Mathematics and Physics, University of Queensland, St Lucia, QLD 4072, Australia, 
$^{2}$School of Mathematical and Physical Sciences, Macquarie University, Sydney, NSW 2109, Australia,
$^{3}$Astrophysics and Space Technologies Research Centre, Macquarie University, Sydney, NSW 2109, Australia,
$^{4}$ Center for Cosmology and Astro-Particle Physics, The Ohio State University, Columbus, OH 43210, USA, 
$^{5}$ Department of Astronomy, The Ohio State University, Columbus, OH 43210, USA,
$^{6}$ Kavli Institute for Particle Astrophysics and Cosmology, Stanford University, 452 Lomita Mall, Stanford, CA 94305, USA
$^{7}$Centre for Gravitational Astrophysics, College of Science, Australian National University, ACT 2601, Australia, 
$^{8}$Research School of Astronomy and Astrophysics, Australian National University, ACT 2601, Australia, 
$^{9}$Sydney Institute for Astronomy, School of Physics, A28, University of Sydney, NSW 2006, Australia,
$^{10}$School of Physics and Astronomy, Monash University, VIC 3800, Australia,
$^{11}$Institute of Cosmology \& Gravitation, University of Portsmouth, Dennis Sciama Building, Burnaby Road, Portsmouth, PO1 3FX, UK
}

\corresp{Hugh McDougall, Email: hughmcdougallemail@gmail.com}

\citeauth{McDougall et al., \red{PLACEHOLDER} {\it Publications of the Astronomical Society of Australia} {\bf 00}, 1--12. https://doi.org/10.1017/pasa.xxxx.xx}

\history{(Received xx xx xxxx; revised xx xx xxxx; accepted xx xx xxxx)}

\begin{abstract}
Reverberation mapping of active galactic nuclei uses the time taken for variability to travel through the system as a measure of system size, with this size being used to inform estimates of the mass of the central black hole, a technique that has graduated to an industrial-scale effort in recent decades. Such large target-count, multi-year surveys have been consistently undercut by poor light curve quality and the aliasing problems of seasonal observations. These impacts compound with statistical approaches that are not consistent between works and sometimes combine inconsistent statistical frameworks. Here, we present a new end-to-end Bayesian framework for fitting lags at scale in large surveys, simultaneous with the fitting of population models such as the radius-luminosity relation, and formulated in a numerically tractable way. This procedure is a statistically and numerically robust generalisation of the stacking and quality cut procedures of past works, fully marginalising over data uncertainty and data quality. Our approach remains numerically stable when scaling to larger surveys, and can act as a basis for any Bayesian population modelling of large sets of AGN time series observations. Applied to the \ozdes sample, we demonstrate that this new approach yields more than double the constraining power on population-level properties like the $R-L$ relationship, and an order-of-magnitude increase in the number of recovered lags and masses. With this method, we go from $24$ high quality lag recoveries with traditional methods to a new effective sample size of $\sim297$, all while being robust against random chance false positive measurements. 
\end{abstract}


\maketitle

\section{Introduction}
\label{sec: intro}
Despite their compact size, the extreme luminosities of active galactic nuclei (AGN) renders them readily observable at cosmological distances. Reverberation mapping (RM) is a technique in which we track the variability of an AGN as it propagates through different elements of the system geometry, using the delay timescale of the propagation as a ruler for the system scale. Broad Line Region (BLR) RM finds the lag between the accretion disk dominated continuum light curves, observed with photometry, and the BLR emission line light curves, isolated with time-series spectroscopy, and uses the lag $\Delta t$, attributed to light travel time between the two regions, to estimate the BLR radius:
\begin{equation}
    R_\mathrm{BLR} \simeq c\cdot\Delta t
    ,
    \label{eq: RM_lag}
\end{equation}
where $c$ is the speed of light. This is commonly used to estimate the mass of the central supermassive black hole (SMBH) by assuming approximate virial equilibrium and balancing the dynamical and gravitational forces:
\begin{equation}    
    M_\mathrm{ BH} = f_\text{vir} \cdot \frac{ R_\mathrm{ BLR} \braket{\sigma_v^2} }{G}, 
    \label{eq: RM_mass}
\end{equation}
where  $\sigma_v$ is the Doppler broadened emission line width, $G$ is the gravitational constant, and $f_\text{vir}$ is the empirically derived virial factor that accounts for uncertainty and population variance in the geometry, kinematics, and alignment of the BLR \citep{Grier_2013b, Woo_2015, Villafana_2023_dynamical, SDSS-Shen_2023, Shen_2024_dynamical}.

By volume of sources, much of modern RM is performed at an industrial-scale, wherein surveys like the Australia Dark Energy Survey \citep[\ozdes;][]{OzDES-DR0-Yuan_2015, OzDES-King_2015} and the Sloan Digital Sky Survey \citep[\sdss;][]{SDSS-Shen_2015} study large samples of AGN over multi-year campaigns that extend over a wide range of redshifts. A major goal in studying such high redshift AGN is to constrain the $R-L$ relationship, a scaling between the luminosity of the AGN and the radius of the broad line region. This is formalised as a power law, or more practically a linear relation on the log-log axis with assumed Gaussian scatter:
\begin{equation}
    \log_{10}(\Delta t_\text{R-L}) = \alpha \log_{10}(\lambda L_{\lambda})+ \beta + \sigma\times\epsilon,
    \label{eq: R-L_powerlaw}
\end{equation}
where $\epsilon$ is a random Gaussian variable such that $\sigma$ is the scatter about the $R-L$ relation. The BLR is believed to be at least partially stratified, \citep{Fausnaugh_2017_stratif} and so different $R-L$ relations are fit for different reverberating broad emission lines. Industrial-scale RM surveys are predominantly in the optical, and so different emission lines are observed from sources at different redshifts: the $4861 \angstrom$ \hbeta for the nearest sources ($z\in \sim[0,0.6]$), the $2798 \angstrom$ \mgii line for intermediate redshifts ($z\in \sim[0.6,1.8]$) and the $1549 \angstrom$ \civ line for the most distant high redshift sources ($z\gtrsim1.8$). These $R-L$ relations allow us to estimate masses for SMBH with only a single epoch of observation, estimating their radii with Equation~\ref{eq: R-L_powerlaw} and using this to estimate mass with Equation~\ref{eq: RM_mass}.

A major issue in multi-year RM is the presence of aliasing: the appearance of false positive lags at \qm{aliasing peaks} at half yearly intervals ($180 \dayu$, $540 \dayu$ etc.). These false positives not only confound our individual SMBH mass measurements, but also contaminate our fitting of the $R-L$ relation. The common approach to the aliasing problem has been a \qm{cut-and-constrain} framework, whereby lags are fit en-masse for an entire survey and then cut or retained by some set of reliability metrics until only the most seemingly reliable measurements are retained for constraining the $R-L$ relations. The field has not reached consensus on exactly how this should be done, with methods of lag measurement and significance benchmarks varying between \ozdes and \sdss and even between individual studies within each survey. 

In past works, \ozdes has used the lag fitting code \javelin and various quality cut measures for each emission line \cite{OzDES-Hoormann_2019, OzDES-Yu_2021, OzDES-Yu_2023, OzDES-Malik_2023, OzDES-Penton_2025}. In \citet[hereafter \paperone]{OzDES-McDougall_2026} we contrasted the existing \ozdes sample with the final data release of \sdss \citep{SDSS-Shen_2023} and a handful of smaller surveys (see that work for details), using the lens of $R-L$ relation constraints to show that the lags reported from \ozdes and \sdss are in tension despite their physically similar target footprint. We showed in that analysis that differences in how lags are measured and what quality cuts are used to gauge their reliability are a major driving factor in differences between surveys, a finding echoed by other works that suggest the need for a more principled and consistent $R-L$ fitting methodology \citep[e.g.][]{Wang_2024_RLhbeta}. For the \mgii sample in particular, differences in methodology between \ozdes and \sdss dominate the tensions between their samples. We also found that failing to account for the ceiling on the recoverable lag had a major impact on the \civ sample's $R-L$ relation, with the inability to properly recover high lags at high luminosity truncating the $R-L$ trend and flattening its apparent slope. 

In \citet[hereafter \papertwo]{McDougall_2025_LITMUS}, we demonstrated that \javelin had a numerical convergence issue in multi-year lag recovery that caused it to greatly magnify the aliasing problem, and presented a new lag fitting code, \litmus, that resolved that issue. \litmus also uses Bayesian model comparison to present principled measures of lag reliability that are consistent with the lag measurements, in place of ad-hoc approaches used in past work. In \citet[hereafter \paperthree]{McDougall_2026_Paperthree} we used \litmus to re-analyse the full \ozdes sample in the cut-and-constrain framework, using Bayes factors as a significance measure and screening potential random chance false positives. We found that:
\begin{enumerate}
    \item The attrition rate of the \ozdes past quality cuts were entirely reasonable for a cut-and-constrain approach, with only $\sim3-10\%$ of the final sample having unmistakably reliable lag measurements.
    \item It is nearly impossible to explain the \mgii sample (and to a lesser extent the \civ sample) as having all of its sources contain recoverable lags, with anywhere from $70-95\%$ of \mgii targets being confidently \qm{lag-free}.
    \item The \ozdes spectroscopic cadence and precision mean that the typical uninformative uniform priors on lag are a major bottleneck on our ability to perform RM at scale.
\end{enumerate}
This third point is worth stressing: modelling in past RM works has adopted an uninformative uniform prior on the lag, and the intentional lack of information from this uniform prior severely limits the number of RM lags we could be confident in, leading to a scarcity of reliable lags for constraining an $R-L$ relation. In \paperthree we showed that the inclusion of a physical prior, i.e. one derived from an $R-L$ relation, had a significant impact on the confidence of our lag recoveries: both improving confidence for true recoveries and rejecting spurious coincidental lags. 
With traditional \qm{cut-and-constrain} approaches to RM informed $R-L$ fitting, we arrive at an impasse: we need the improved constraining power of an $R-L$ prior to recover lags, but we cannot use the resulting lags to fit an $R-L$ relation without inherent confirmation bias towards whatever existing $R-L$ relation we made use of.

In this work we adopt a new statistical framework for fitting the $R-L$ relationship and lags simultaneously that fully addresses all of the shortfalls and complications of industrial-scale RM and $R-L$ fitting. This hierarchical model is end-to-end Bayesian and makes use of the entire set of available information from every \ozdes AGN target. It differs from past work in four ways:
    \begin{itemize}
        \item Every source is fit instead of retaining only a few \qm{high quality} sources,
        \item We use the full posterior for each lag source instead of lossy summary statistics,
        \item We allow for, and estimate the rate of, AGN that do not contain recoverable lags,
        \item We fully account for the lag ceiling and observational windowing function, and 
        \item We fully marginalise over uncertainty and issues of data quality.
    \end{itemize}
This hierarchical model is simultaneously a generalisation of the weighted $R-L$ fitting procedure \citep{OzDES-Penton_2025}, and also of the stacking technique in which information is combined from multiple physically similar targets \citep[e.g.][]{Fine_2012_stacking, Fine_2013_stacking, Li_2017_stacking, OzDES-Malik_2024, Aalfarsy_2026_stacking}. Within AGN RM, such hierarchical models have been examined in a limited basis elsewhere \citep{Brewer_Elliot_2014_hierarchical}, but we present here a means of calculating them that is numerically tractable and easily scalable for nearly arbitrary sample sizes and model complexity. For a given model of AGN variability and lag response, and for a given model of the AGN population, we suggest that the framework we present here may act as the basis for the most complete approach for performing AGN RM at scale.

We have structured this paper such that the main body presents the general structure of our new method and demonstrates its results and efficacy on real data, while the appendices contain more specific details on numerical and statistical considerations for the reader interested in altering or extending this work. In Section~\ref{sec: Data} we discuss the \ozdes data that we apply our method to. In Section~\ref{sec: RM_stats} we provide an overview of the statistical methods and modelling needed for Bayesian RM in general, along with a summary of the methods used in \paperthree which act as inputs to our hierarchical model. In Section~\ref{sec: methods} we describe our new hierarchical model and its numerical basis, supplemented by details in Appendix~\ref{app: num}. We apply our method to the \ozdes sample in Section~\ref{sec: results}, and demonstrate that, compared to traditional methods, it yields and order-of-magnitude increase in the number of measured lags and SMBH masses and improves $R-L$ constraining power by multiple times. In Section~\ref{sec: widerimpact} we examine this framework in the broader context of AGN RM, including how it may be applied to upcoming or ongoing RM campaigns like \desi \citep[e.g.][]{Aalfarsy_2026_stacking} and \tides \citep{Frohmaier_2025_4mostRM}, and we conclude in Section~\ref{sec: conclusion} by providing suggestions for future applications for and refinements of our method.

\section{The \ozdes Sample}
\label{sec: Data}
To demonstrate our method in this work and its improvement over traditional methods, we make use of the same data set as used in \paperthree. AGN light curves are drawn from the $7$-year \ozdes project, \citep{OzDES-DR0-Yuan_2015,OZDES-DR1-Childress_2017,OZDES-DR2-Lidman_2020} which provided spectroscopic follow up to the photometric imaging of the Dark Energy Survey \citep[\des;][]{DES_2016_REF}. \ozdes was performed using the AAOmega spectrograph fed by the Two Degree Field (2dF) fibre positioner on the 3.9m Anglo-Australian Telescope at Siding Spring Observatory, \citep{Lewis_2002_2df, OzDES-DR0-Yuan_2015, OZDES-DR1-Childress_2017} offering a spectral resolution of $R=1400$ to $1700$ in the wavelength range of $3700\,\angstrom$\ to $8800\,\angstrom$, while the \des \decam measurements were performed on the CTIO Blanco 4-metre Telescope in Chile.

Our photometric measurements were made in the $g, r, i, z$ filters with approximately weekly cadence, while spectroscopic light curves were sampled roughly once a month, with both samples having $5-6$ month observational seasons each year due to the sun. Where we refer to the continuum measurements in this work, we follow convention set by other \ozdes papers and use the $g$-band photometry from \des \citep{Flaugher_2015}, cleaned of outlier observations using the method described in Appendix~A of \paperthree. 

Our emission line response light curves and spectra are produced using the pipeline of \citet{OzDES-Hoormann_2019} for the \hbeta and \civ sources, and that of \citet{OzDES-Yu_2021} for \mgii. In this work we are limited to only AGN with readily measured luminosities and well structured continuum light curve variability\footnote{well structured light curves being those with high $\BF{Struc,Const}$ per Section~\ref{sec: RM_stats_BF}}, which is a total of $797$ sources, comprised of $69$ \hbeta sources, $448$ \mgii sources and $280$ \civ sources. Where measuring SMBH masses with Equation~\ref{eq: RM_mass}, $2$ \hbeta sources, $8$ \mgii sources and $23$ \civ sources do not have line widths available, and are excluded from our mass sample but not our $R-L$ fitting. For the full set of pre-cleaning light curves, luminosities and velocities, see the data release accompanying the \ozdes reverberation mapping data release in \paperone, while the post-cleaning continuum light curves are in the data release of \paperthree.

\section{Background Statistical Modelling}
\label{sec: RM_stats}
In this section we summarise the statistical concepts that we employ in our hierarchical RM lag model. In Section~\ref{sec: RM_stats_LCmodel} we discuss the GP based lag recovery models used in prior RM works, and in Section~\ref{sec: RM_stats_BF} we describe the Bayesian hypothesis testing and model comparison measures used in \paperthree.
The reader familiar with HM3 can skip to  Section~\ref{sec: methods}, where we introduce our Bayesian hierarchical model for RM, as well as the Monte Carlo integration scheme that we use to make these high dimensional problems tractable.

\subsection{Lag Constraints \& Light Curve Modelling}
\label{sec: RM_stats_LCmodel}
In \paperthree we describe the particulars of our light curve modelling in detail, and we present them briefly here. The core idea of all RM techniques is to assume the observed broadband continuum emission light curve, $y_c(t)$, presents some echo at time $\Delta t$ in the response light curve, $y_r(t)$. If we allow the continuum and response to have some noise process $n_{r/c}$ such that the echo is apparent only in some underlying signal $s_{r/c}(t)$, this can be formalised as:
\begin{eqnarray}
    \label{eq: RM_signalnoise}
    y_c(t) &=& s_c(t)+n_c(t)\\ \nonumber
    y_r(t) &=& s_r(t)+n_r(t)\\ \nonumber
    s_r(t) &=& s_c(t)\star\psi(t-\Delta t)
    ,
\end{eqnarray}
where $\star$ represents a time-domain convolution and $\psi(t)$ is a transfer function, centred at lag time $\Delta t$, which describes the scaling and smoothing of the response as compared to the continuum. In \paperthree we refer to any model that can be described in this way with a unimodal transfer function as \qm{call and response} model. GP-based methods are those in which $s_c(t)$, and by extension $s_r(t)$, can be described as a stochastic process with a power spectral density (PSD) of a known functional form. In such GP models, the continuum is characterised by some autocovariance kernel, $\phi_{cc}(t,t') = \braket{y_c(t),y_c(t')}$, usually assumed to be stationary such that $\phi_{cc}(t,t')=\phi_{cc}(\delta t), \ \delta t=t'-t$. The covariance between the signal and response, $\phi_{rc}$, is found by convolving this kernel with the transfer function, and the response autocovariance, $\phi_{rr}$, is arrived at by convolving once more, i.e.:
\begin{eqnarray}
    \phi_{rc}(\delta t)
    &=& \int_{-\infty}^{\infty}{\phi_{cc}(\delta t)\psi(\delta t-t')}dt'
    \\
    \phi_{rr}(\delta t)&=& \int_{-\infty}^{\infty}\int_{-\infty}^{\infty}{\phi_{cc}(\delta t)\psi(\delta t-t')\psi(\delta t-t'')}dt'dt''
    .
    \label{eq: conv_transfers}
\end{eqnarray}
With these covariance functions, we can construct a covariance matrix, $C=S+N$, where $S$ represents the reverberating components of the signal structure for both continuum and response, while $N$ encodes the noise processes. In the simplest form we attribute the noise purely to measurement uncertainty, i.e.:
\begin{equation}
    C = S + N, \; 
    \begin{array}l
    N_{ij}=\delta_{ij} E_i E_j\\
    S_{ij}=\braket{y_i,y_j}
    \end{array}
    ,
    \label{eq: covar_matrix}
\end{equation}
where $E_{i}$ are the measurement uncertainty in light curve signal observations $y_{i}$ and $y_{j}$ and $\delta_{ij}$ is the Kronecker delta such that $N$ is diagonal. Equipped with this matrix, the likelihood for a vector of observations $\vec{y}$ (after subtracting off the continuum and response means) is a simple multivariate Gaussian:
\begin{equation}
    \mathcal{L}(d^n \vert \theta^n) = \frac{1}{\sqrt{(2\pi)^N\det(C)}} 
    \exp\left(-\frac{1}{2} \vec{y}^TC^{-1}\vec{y} \right)
    ,
    \label{eq: GP_likelihood}
\end{equation}
where $\theta$ are the all model parameters that go into constructing $C$ and $\vec{y}$. This equation is the core of all GP-based fitters: different light curve models rely on varying descriptions of the signal mean, the GP kernel, and in their description of the noise processes. In this work, as in other \ozdes analyses, we treat the signal mean as being stationary for both continuum and response, and the noise as being uncorrelated white noise. Non-diagonal noise processes (e.g. structured but non-reverberating variability in either light curve) may be expressed with a non-diagonal $N$.

It is common to model AGN light curves as a Damped Random Walk (DRW), i.e. as a red noise process that flattens out to stationarity at timescales much longer than some characteristic timescale $\tau$ \citep{Kelly_2009, Kozlowski_2010a, MacLeod_2010}. In a DRW, the continuum is described by a symmetrical exponential function:
\begin{equation}
    \phi_{cc}(t_i,t_j)=\sigma_c^2 \exp \left( -\abs{\frac{t_i-t_j}{\tau}} \right)
    .
    \label{eq: DRW_covar}
\end{equation}
In the simplest model, what we refer to in \papertwo and \paperthree as the \qm{Simple DRW} model, we ignore any non-measurement noise processes as well as any smoothing of the response light curve. In such a model, the transfer function is a simple shifting and scaling of the continuum, $\psi=\frac{\sigma_r}{\sigma_c}\delta(t-\Delta t)$, $\delta t$ here being the Dirac delta. 

In codes like \javelin, a degree of smoothing is permitted between the continuum and response, encoded in a \qm{tophat} transfer function with a smoothing timescale of $w$ days:
\begin{equation}
    \psi(t) = \frac{1}{h}\left\{\begin{array}{lr}
        1 , \; \Delta t-\frac{w}{2}<t<\Delta t+\frac{w}{2} \\
        0 , \; \mathrm{otherwise}.
    \end{array}\right.
    .
    \label{eq: tophat}
\end{equation}
In \paperthree, we describe such a model as the \qm{DRW + Tophat} model. With such a transfer function, the line-response covariance function becomes
\begin{multline}
    \phi_{rc}(t_i,t_j)
    = \sigma_r \sigma_c\frac{1-\exp \left(-\frac{w}{2\tau}\right)}{\sqrt{\frac{1}{2}(\frac{w}{2}+\exp\left(-\frac{w}{2}\right)-1)}}
    \times \\
    \left\{\begin{array}{lr}
        1 - {(
            \cosh\left(\frac{\delta u^{ij}}{\tau}\right) - 1
        )}
        {(
            \exp(\frac{w}{2\tau})-1
        )^{-1}} \
        , \;
        \delta u^{ij} < \frac{w}{2} \\
        \exp\left(\frac{\delta u^{ij}}{\tau}\right)
        {(
            \sinh(\frac{w}{2\tau})
        )}{(
            \exp(-\frac{w}{2\tau})
        )^{-1}}
        , \;
        \delta u^{ij} > \frac{w}{2}
    \end{array}\right.
    \label{eq: rc_covar},
\end{multline}
where, for continuum and response measurements at time $t_i$ and $t_j$, $\delta u^{ij} = \lvert t_i-t_j-\Delta t \rvert $\footnote{Note that this centres the response at the middle of the tophat, corresponding to the \qm{tophat-centroid} model of \paperthree}. The response's self covariance is, resulting from a double-convolution per Equation~\ref{eq: conv_transfers},
\begin{multline}
    \phi_{rr}(t_i,t_j) =\sigma_r^2 \times \\
    \left\{\begin{array}{lr}
        \frac{w-\delta t^{ij}}{\tau} + \exp(-\frac{w}{\tau}) \times \cosh(\frac{\delta t^{ij}}{\tau}) - \exp(\frac{-\delta t}{\tau})
        , \; \delta t^{ij} < w\\
        \exp(\frac{-x}{\tau})  \cosh(\frac{w}{\tau})-1
        , \; \delta t^{ij} > w
        \end{array}\right.
    \label{eq: rr_covar},
\end{multline}
where $\delta t^{ij}=\abs{t_i-t_j}$.

In higher cadence or signal-to-noise ratio (SNR) sources, there has been work in examining more expressive transfer functions to encode the specifics of the BLR response, for example the multi-peaked Gaussians of \citet{Li_2016_MICA}. In \paperthree, we found that the sparse cadence and wide measurement uncertainty of \ozdes meant that the exact choice of transfer function did not have a marked impact on lag recovery.

\begin{figure}
    \centering
    \includegraphics[width=1.0\linewidth]{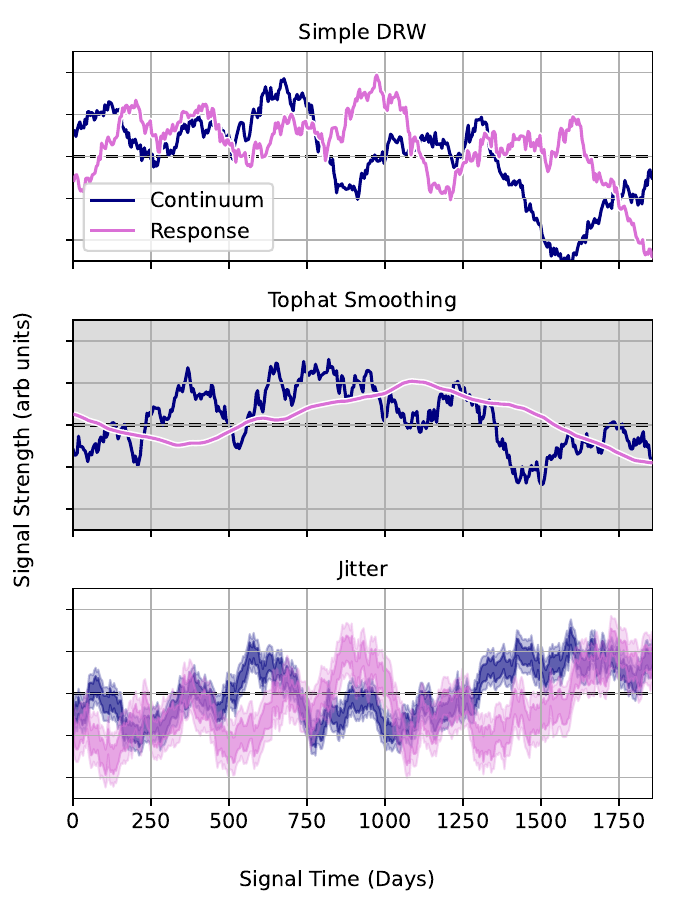}
    \caption{Examples of simulated light curves for our three light curve models, for both continuum (navy) and response (purple), set at a timescale of $\tau=400 \dayu$ and a lag of $\Delta t=300 \dayu$. From top to bottom we have examples of the simple DRW with no smoothing or jitter, tophat smoothing of the response per Equation~\ref{eq: tophat}~-\ref{eq: rr_covar} with a smoothing timescale of $b=400\dayu$, and jitter on both continuum and response light curves. The tophat smoothing model is shaded grey to indicate its use as our fiducial model.}
    \label{fig: lc_examples}
\end{figure}

In \paperthree we also examined the effect of allowing for additional noise in the continuum and response signals, a simple first order correction by adding a source of additional variance to $n_{c/r}(t)$ that we referred to as a \qm{jitter} model. To keep the total signal variance constant between models, we construct this jitter on a sliding scale from $0-100\%$, with zero being a pure DRW and $100\%$ being pure unstructured white noise. Under such a model, we adjust the signal covariance matrix to:
\begin{equation}
    S_{ij} = S_{ij,\text{DRW}}\sqrt{1-\delta_{ij}J_iJ_j}+\delta_{ji}J_{i}J_{j}\sigma_i\sigma_j,
    \label{eq: jitter_covar}
\end{equation}
where $S_{ij,\text{DRW}}$ is the simple DRW model signal covariance from Equation~\ref{eq: covar_matrix}. This jitter model is a catch-all for any way that the simple DRW model fails: additional astrophysical variance in the signal above that of a DRW, under-estimates of the calibration error in the construction of the light curves, etc. Example light curves demonstrating these three models of the signal and noise processes of the continuum and response are shown in Figure~\ref{fig: lc_examples}. In this work, we use the tophat-smoothed response with no jitter as our fiducial model, but test all three light curve in parallel. Their relative performance is discussed in Section~\ref{sec: widerimpact_modelling_LCmodels}.

\subsubsection{Null Hypothesis \& Lag Recovery Significance}
\label{sec: RM_stats_BF}
In \paperthree we described four model hypotheses that could describe the paired light curves for a given AGN source, each described by a different Bayesian model. After selecting a light-curve model above, any pair of continuum and response light curves may be described by one of four possibilities:
\begin{enumerate} 
    \item The \qm{coupled} model in which the light curves are both GPs with some observable lag $\Delta t$; 
    \item An \qm{uncoupled} model, in which both light curves are well described by a GP but with either complete decoupling or a lag so long it appears as if they are decoupled;
    \item A \qm{GP-Noise} model, in which the continuum is well structured as a DRW but the response is completely stationary (i.e. constant) or unstructured (i.e. white noise);
    \item A \qm{Noise-Noise} model in which both the continuum and noise are unstructured.
\end{enumerate}
Under each of these models, we can evaluate an \qm{evidence} or \qm{marginal likelihood}
\begin{equation}
    \mathcal{L}(d\vert\mathrm{Model})=\ev_\text{Model}(D) = \int \pi_\text{Model}(\theta)\mathcal{L}_\text{Model}(d\vert\theta) d\theta
    .
    \label{eq: evidence_definition}
\end{equation}
where $\pi(\theta)$ represents the prior distribution for all model parameters, both the lag $\Delta t$ and the nuisance parameters. By itself the evidence tells us little, so we instead use the Bayes factor between two models, i.e. the evidence ratio,  ($\BF{a-b}=\frac{\ev_a}{\ev_b}$), to determine which hypothesis is best supported by the data. In \paperthree we provided three such Bayes factors when fitting for lag under uniform lag priors:
\begin{itemize}
    \item $\BF{Lag,\fidp}$ comparing the coupled ($1$) and uncoupled ($2$) models and telling us about the strength of the lag detection, i.e. whether the apparent alignment of the continuum and response at some lag could be produced by random chance with two stochastic signals,
    \item $\BF{Struc,Resp}$ comparing the uncoupled ($2$) and GP-noise ($3$) models and telling us about how much structured variability is observed in the response light curve, and
    \item $\BF{Struc,Cont}$ comparing the GP-noise ($3$) and noise-noise ($4$) models and telling us how much structured variability is seen in the continuum light curve.
\end{itemize}
Note that in our notation here subscript $\BF{Lag,\fidp}$ with $\fidp$ to indicate that it is the Bayes factor when fitting with a fiducial uniform prior on $\Delta t$. In \paperthree we truncated our Bayesian analysis at this stage, using these Bayes factors as measures of significance in a frequentist test-statistics sense and comparing false positive rates to a set of lag-free light curves. In this work, we have no such truncation: a Bayesian hierarchical model is used to fit all lags for even sources with weakly recovered lags simultaneous with the AGN population properties.

\subsection{Fiducial Lag Fits \& Nested Sampling}
\label{sec: RM_stats_stats_nestedsampling}
Most RM works have concerned themselves with only constraining the lag, in the specific context of Bayesian works this means constraining the shape of the lag posterior distribution under broad priors, i.e. exploring the shape of the likelihood in Equation~\ref{eq: GP_likelihood}. In \papertwo and \paperthree we demonstrated the importance of not just the shape of the posterior distribution, but also its normalisation, i.e. the Bayesian evidence in Equation~\ref{eq: evidence_definition}. RM fitters like \javelin and \mica use Markov Chain Monte Carlo Methods, which evaluate the shape alone. In simple terms, they evaluate the \qm{best fit} lag, but offer no comment as to whether this represents a \qm{good fit}. Past works have instead combined Bayesian GP-based lag measurements with a patchwork of lag significance measurements from different statistical methods (e.g. linear interpolation of the light curves) or post-hoc analysis of the posterior distribution shape (see \paperone Section~$4.2$ for a summary). In \paperone we demonstrated that these choices were inconsistent and unreliable, and in \papertwo we showed that the lag-no lag Bayes factor ($\BF{Lag,\emptyset}$) offers an effective and statistically principled means of weighing the significance of a particular lag recovery in a framework consistent with the lag measurement itself. To calculate such evidences, we need to be able to integrate the posterior distributions of our lag fits per Equation~\ref{eq: evidence_definition}.

Nested Sampling \citep[NS;][]{Skilling_2006_NS} is a robust multidimensional integration scheme well suited to evidence integrals that can also be used to generate posterior samples. In \papertwo we identified that the likelihood function of seasonal light curves exhibit a multimodality \cite[also explored in][]{Kroupa_2026b_likelihood} that is both pathological to many Markov Chain Monte Carlo (MCMC) fitters \citep[e.g. the fitter \emcee by][as used by \javelin]{emcee-Foreman_Mackey_2013}. Many nested sampling implementations, though often computationally slower and less scalable than MCMC methods, are robust against this multimodality, although nested sampling can have difficulties in extremely high SNR RM (see \citet{Kroupa_2026b_likelihood}). NS is also distinct from MCMC-like methods in that it samples the entire domain, rather than MCMC methods which sample only the high probability regions of parameter space. We provide here only the details relevant to this work, but suggest a good overview in \citet{Speagle_2020_dynesty} for details on particular implementations.

NS in general operates by starting with a set of sample points spanning the entire prior domain, and then iteratively drawing this ensemble inwards towards the posterior peaks in a process of \qm{shrinkage}, with each iteration yielding a single nested sample of known contribution to the final evidence integral. By the end of an NS run, we have an ordered set of points $\{\theta_j\}$, each of known likelihood $\mathcal{L}_j$, posterior density $\rho_j$, volume $V_j$ and evidence contribution $\Delta \ev_j$. These nested samples differ from posterior samples in three main respects: (1) unlike posterior samples, they span the entire domain of the parameter space, (2) they are of unequal posterior weighting and (3) unlike posterior samples they have a strict ordering and cannot be randomly re-ordered or sub-sampled. Nested samples may be reprocessed into posterior samples by randomly drawing them with a draw weight $w_j=\frac{\Delta \ev_j}{\ev}$, i.e.:
\begin{equation}
    \theta_{\text{Post},i}
    \sim
    \mathrm{Redraw}\left(
    \theta_{\text{Nest.}j},\left(\frac{\Delta \ev_j}{\ev}\right)
    \right)
    .
\end{equation}

In \paperthree we used NS, specifically fit with \jaxns, \citep{Albert_2020_JAXNS} for our lag fitting on a source-by-source basis with uninformative uniform priors on the lag. In this work, we have re-run those fits with uniform priors up to $\tmax$, using the same arbitrarily broad priors on the non-lag parameters discussed in \paperthree, and use these fiducial pre-fits as inputs to the hierarchical model we discuss in Section~\ref{sec: methods_practical_calc}. When running these fits we save the full chain of nested samples rather than the re-weight derived posterior samples for reasons of numerical stability, discussed in detail in Appendix~\ref{app: num_nestsampstability}.

\section{A New Hierarchical / Stacked $R-L$ Model}
\label{sec: methods}

\begin{figure*}
    \centering
    \includegraphics[width=1.0\linewidth]{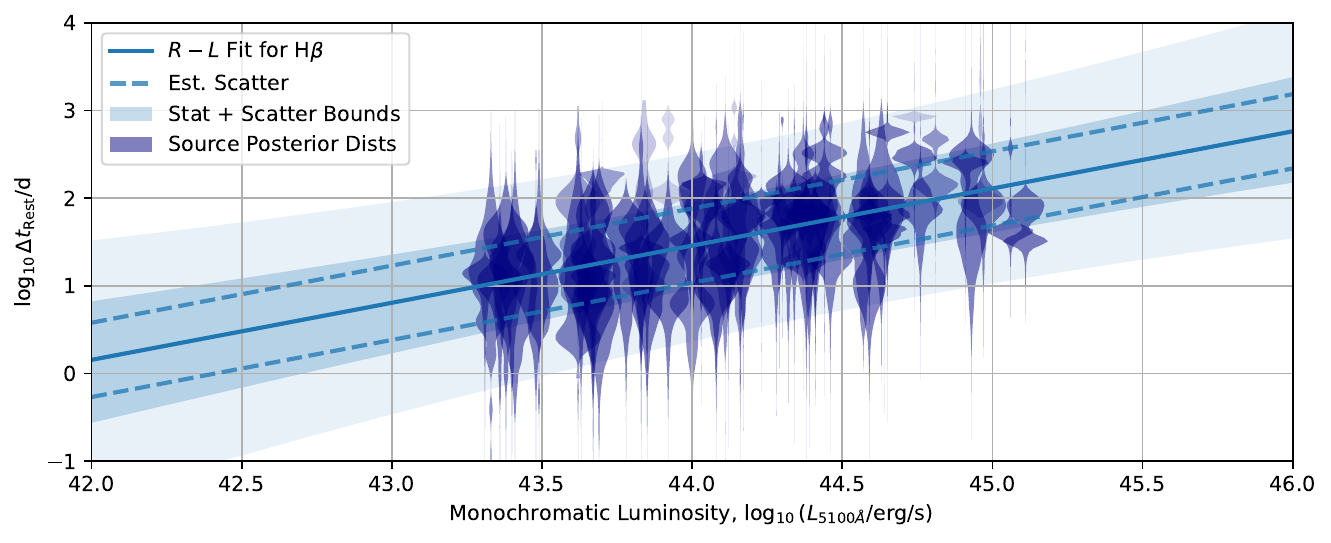}\\
    \includegraphics[width=1.0\linewidth]{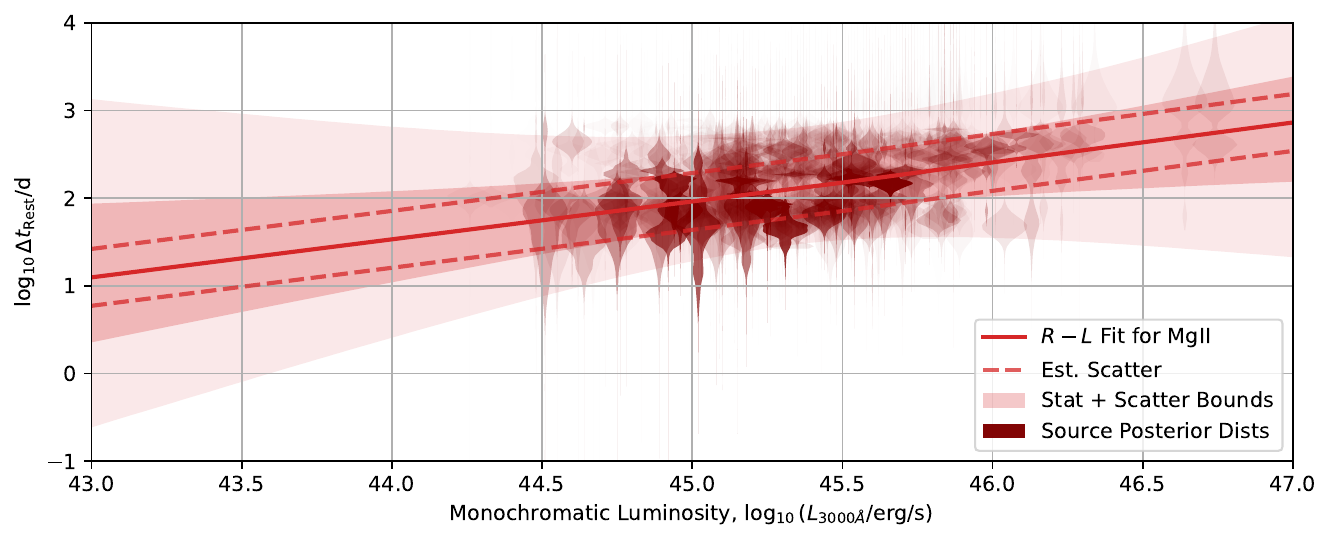}\\
    \includegraphics[width=1.0\linewidth]{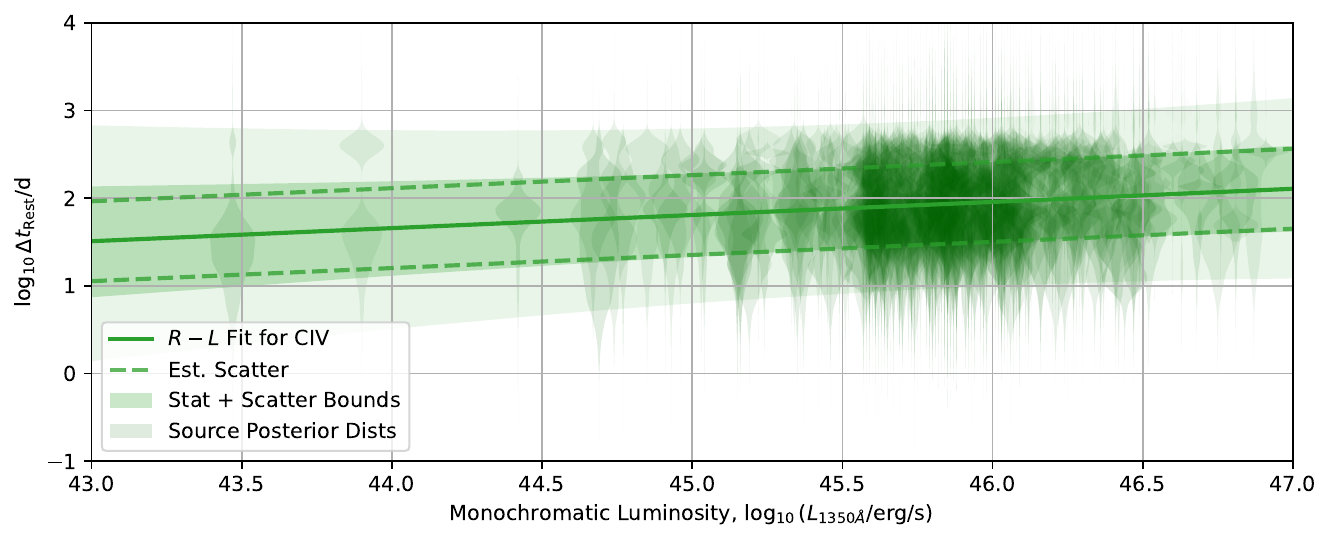}
    \caption{A demonstration of our recovered radius luminosity relationship and the constraints on individual lags for each source in our sample after marginalising over our uncertainties in the $R-L$ relationship parameters. Sources are opacity-shaded by their marginalised probability of containing a lag $P^n_\mathrm{Lag}$ per equation~\ref{eq: plag}, and the posterior distributions for $\Delta t_\mathrm{Rest}$, marginalised over the posterior distribution for the population $R-L$ parameters, are shown with violin plots centred at the source luminosities. The low and high opacity shaded bands represent the 1 and 2 sigma uncertainties, inclusive of statistical uncertainty. The dashed lines show the posterior median population scatter about the posterior median prediction, indicating the minimum scatter we could get in a perfectly constrained $R-L$ relation.}
    \label{fig: RL_violin}
\end{figure*}

\begin{figure}
    \centering
    \includegraphics[width=0.6\linewidth, angle=0, trim={0.0cm 3.8cm 21.3cm 4cm}, clip]{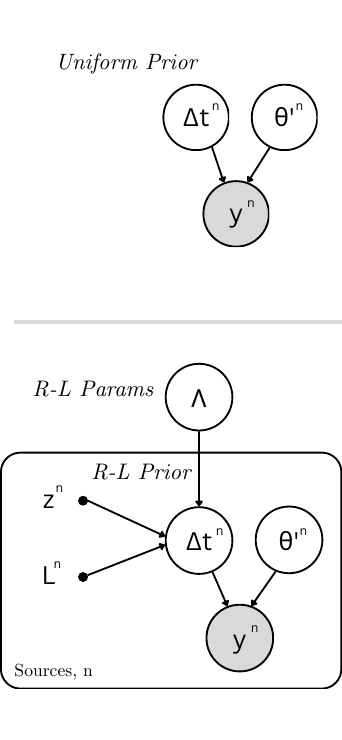}
    \caption{Probabilistic graphical models showing the Bayesian models for the typical source-by-source model of lag recovery (top panel) and the hierarchical model (bottom panel). In the traditional source-by-source framework, the lag ($\Delta t$) and non-lag parameters ($\theta^\prime$), both fit with uninformative uniform priors, are inputs to the forward model for the observed light curves ($y$) for each source in isolation. For the hierarchical model, the population parameters ($\Lambda$), in concert with source luminosity ($L$) and redshift $z$, offer an informative log-normal prior on the lags. Fit jointly, we gain information about the lags from the population, and the population from the lags.}
    \label{fig: prob_graphs}
\end{figure}

\begin{figure*}
    \centering
    \includegraphics[width=1.0\linewidth, angle=0, trim={6.5cm 3.8cm 0.4cm 2cm}, clip]{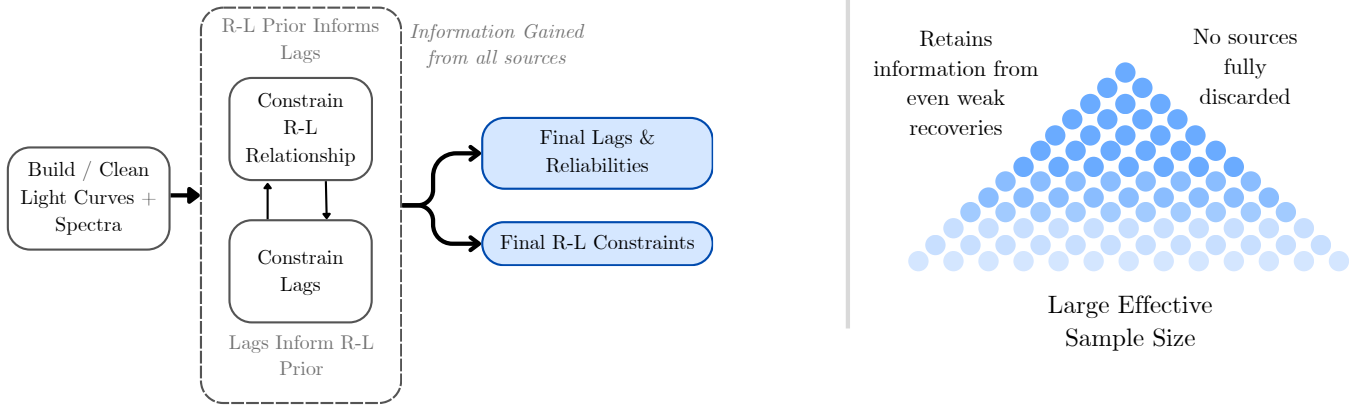}
    \caption{Sketches indicating the difference between the traditional cut-and-constrain method of lag measurement / $R-L$ constraints (top panel) and the fully hierarchical method we propose here. The left panels in each row shows a Bayesian probabilistic graphical model, while the right panel shows a conceptual flowchart of the process. In the traditional approach, lags are fit with uniform priors one at a time, with the quality of sources then being estimated by some measure of merit (e.g. the negative lag tests of \citet{CIV_Grier_2019} and \citet{SDSS-Shen_2023}) and poorly performing measurements being \qm{cut}, with the retained lags being used to \qm{constrain} the $R-L$ relationship. In the hierarchical model, the $R-L$ prior and its parameters, $\Lambda$, are fit simultaneously with the lags. In this approach, no lags are fully discarded, and they all benefit from the information supplied by the $R-L$ informed prior.}
    \label{fig: schematic_example}
\end{figure*}

Shortfalls of the existing cut-and-constrain approach are twofold: firstly, it discards the overwhelming majority of lag recoveries and secondly it fits without making use of our understanding of the AGN population. There is a solution to both of these limitations: a Bayesian hierarchical model in which we fit the lags using a prior informed by the $R-L$ relationship simultaneous with using these lags to fit the parameters that describe this prior (we show the marginalised results of such a model in Figure~\ref{fig: RL_violin} and contrast it with the source-by-source modelling in Figure~\ref{fig: prob_graphs}). Figure~\ref{fig: schematic_example} demonstrates the difference between these two methods: in the cut-and-constrain approach we assume uninformative (uniform) priors for $\Delta t$, $\pi_\emptyset(\Delta t)$, and fit each lag individually. We then use some quality measures to retain only the most reliable of these measurements, and use these to put constraints on the $R-L$ relationship parameters. In the hierarchical model, each source $n$ has a prior on lag  informed by the population at large, $\pi^n(\Delta t \vert \Lambda)$, where $\Lambda$ are the hyper-parameters describing the entire population of sources (e.g. the slope, offset and scatter of the $R-L$ relationship). In the hierarchical model, the likelihood is a joint distribution of all parameters over all sources along with these population properties:
\begin{eqnarray}
    \label{eq: hierarchical_probability}
    \mathcal{L}(D\vert\Lambda,\{\Delta t^n, \theta^{\prime n}\})=&
    \pi(\Lambda \vert D)
    \prod_{n}
    \sum_\mathrm{Model}
        \pi(\Delta t^n\vert\Lambda)\\ \nonumber
         &\times\pi(\theta^{\prime n}) \mathcal{L}_\mathrm{Model}(d^n \vert \Delta t^n,\theta^{\prime n})    
    .
\end{eqnarray}
Here, $D=\{d^n\}$ represents the totality of all observations from each source, and $d^n$ is the observed data only for source $n$, while the likelihood is that of a particular light curve model as described in Section~\ref{sec: RM_stats_LCmodel}.

\begin{figure}
    \centering
    \includegraphics[width=1.0\linewidth]{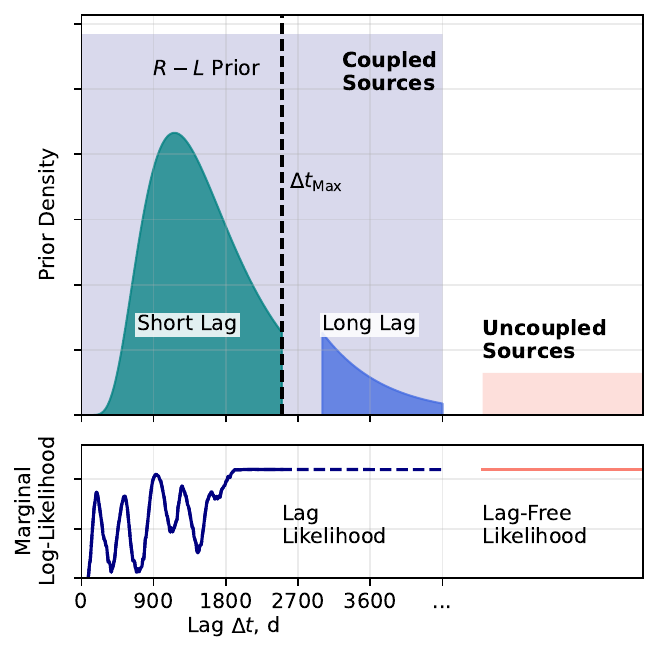}
    \caption{An illustration of how we share the prior and likelihood between the coupled and uncoupled models. For $\Delta t > \tmax$, we use the same lag-likelihood as the uncoupled model, but using the prior contribution from the coupled model.}
    \label{fig: RLprior_example}
\end{figure}

In this work we offer four models to explain any given source, slightly adjusting the models from our source-by-source modelling. These models are:
\begin{enumerate} 
    \item \label{mod: 1} The coupled model in which the signals encode a lag, $\Delta t$, with either:
    \begin{enumerate}
        \item \label{mod: 1a} This lag being possible to observe, i.e. below some limit $\Delta t_\mathrm{Max}$, or
        \item \label{mod: 1b} The lag being too long to observe with our survey baseline, such that the signals appear uncoupled.
    \end{enumerate}
    \item \label{mod: 2} The \qm{uncoupled} model, in which both light curves are well structured (well explained by a GP) but are uncoupled, and
    \item \label{mod: 3} The \qm{GP-Noise} model, in which the continuum is well structured but the response is stationary.
\end{enumerate}
In uniform prior modelling, there is no distinction between model~\ref{mod: 1b} and \ref{mod: 2}: if the lag between continuum and response is too long it looks the same as if they are completely decoupled. In this work, using constrained $R-L$ informed priors, there is a distinction: model~\ref{mod: 1b} shares a likelihood with \ref{mod: 2}, but shares a prior with model~\ref{mod: 1a}. We illustrate this in Figure~\ref{fig: RLprior_example}.

Note that we exclude the \qm{Noise-Noise} model, i.e. that in which the continuum light curve is unstructured as we found in \paperthree that all but 4 sources have clear structure in their continuum light curves ($\BF{Struct,Const}>100$). The four sources without good structure are simply excluded from our sample here.

We are concerned with putting constraints on $\Lambda$, namely finding its posterior distribution $P(\Lambda\vert D)\propto\pi(\Lambda)\mathcal{L}(D\vert\Lambda)$. We have to contend with the fact that our modelling allows that some sources do not have a recoverable lag. As such, the marginal likelihood for any source has two components, one for the source having a lag and one for without:
\begin{eqnarray}
    \ev^n_\mathrm{Coup}(d^n\vert\Lambda)
    &=&
    \int \pi(\Delta t^n \vert \Lambda)\mathcal{L}_\text{Coup}(d^n\vert \Delta t^n) d\Delta t^n
    .
    \label{eq: component_evidences}
\end{eqnarray}
Note that only $\mathcal{Z}^n_{\text{Coup}}$ depends on $\Lambda$. This is because $\Lambda$ only informs the prior for the lag, $\Delta t$, which is not a parameter in the lag-free component. In this work, we use a log-normal prior for lags informed by a simple $R-L$ relationship:
\begin{align}
    \pi^n(\Delta t\vert\Lambda) =
    \frac{1}{\ln(10)\Delta t}
    &\times
    \mathcal{N}(\log_{10}(\Delta t) \vert \mu_{R-L}^n, 
    \sigma)
    ),
    \label{eq: R-L_prior}
\end{align}
where $\mu_{R-L}^n(\Lambda)=\beta + \alpha(L^n_\lambda-N)+\log_{10}(1+z^n)$ is the median rest-frame lag predicted by the $R-L$ relation at parameters $\Lambda$ for source $n$ at luminosity $L^n_\lambda$ and redshift $z^n$. We discuss this further in Section~\ref{sec: widerimpact_modelling_popmodels}, but more complex population models can also be fit by altering the functional form and arguments of this prior.

We can now break the marginal likelihood for an event with the $R-L$ informed prior $\pi^n\left(\Delta t \vert \Lambda\right)$ into the coupled and uncoupled model components:
\begin{eqnarray}
    \label{eq: marg_lik_all}
    \mathcal{L}\left(D\middle \vert \Lambda,f\right)
     & = & 
     \prod_{n}{\mathcal{L}\left\{d^n\middle \vert \Lambda\right)} 
     \\ \nonumber
     & = & 
     \prod_{n}\sum_\text{model} \pi(\text{Model}) \mathcal{L}_\text{Model}\left(d^n\middle \vert \Lambda\right) 
     \\ \nonumber
     & = & \prod_{n}{
     \left[
     f\cdot \ev^n_{\text{Coup}}\left(d^n\vert\Lambda\right) + \left(1-f\right) \cdot \mathcal{Z}^n_{\text{Uncoup}}(d^n)
     \right]} 
     .
\end{eqnarray}
It is worth considering the meaning of the $f$ term here. In \paperthree, this represents a lower bound on the number of AGN with clear lags below some upper limit on lag. In this work, it represents the prior probability that any given AGN hosts a call and response lag. By including this term and allowing the model to fit sources as being either coupled or uncoupled, we allow the fitting to naturally filter out false positives or poorly performing lag recoveries and down weight weak recoveries as a part of the fitting. This gives insight into the physical sample properties, but also marginalises out data issues of quality and lag significance. We show in Figure~\ref{fig: RL_violin} a demonstration of this model as applied to the \ozdes sample for all lines, with the posterior predicted $R-L$ relation bands shown along with the marginal distributions for all lags in the sample (results discussed in detail in Section~\ref{sec: results_RL_constraints}). In this figure, we can see how the $R-L$ relation informs the lag constraints for poor SNR sources, but also tracks the distribution of the lags over luminosity.

\subsection{A Practical Calculation Method via Importance Sampling}
\label{sec: methods_practical_calc}
Fitting the posterior distribution for Equation~\ref{eq: marg_lik_all} is numerically intractable by most methods if attacked directly, as jointly fitting all parameters of all sources simultaneously (e.g. integrating over source parameters in Equation~\ref{eq: component_evidences} with MCMC) would mean exploring a parameter space of hundreds to thousands of dimensions. There exist high dimensional Bayesian fitters that work to this order (e.g. stochastic variational inference \citep[SVI;][]{Hoffman_2013_SVI} or Hamiltonian Monte Carlo \citep[HMC;][]{Neal_1996_HMC02}), but these are unsuited to the complex multi-modal distributions of the lag likelihood function (see \papertwo for an exploration of these challenges and their sources).

Here, we make use of the technique of Monte Carlo integration via importance sampling to simplify this to a low dimensional and \sqm{well behaved} numerical problem in which we directly fit only the population parameters. The rationale is this: there is no need to explore every dimension of this complex multivariate problem when we already have information about the shape of the distribution from our fiducial source-by-source fits. By \qm{recycling} samples from these fiducial pre-fits, we can numerically pave over the complexity of the problem and render it a simple, approachable problem.

First, we note that we have some set of samples from posterior or nested samples for the lag posterior under the uninformative uniform prior up to a bound of $\tmax$, along with the lag-bearing model evidence $\ev_{\mathrm{Lag},\emptyset}$, 
lag-free model evidence $\ev_{\mathrm{NoLag}}$ and Bayes factor $\mathrm{BF}_{\mathrm{Lag},\emptyset}=\frac{\ev_{\mathrm{Lag},\emptyset}}{\ev_{\mathrm{NoLag}}}$. We then say that the prior-informed lag-bearing evidence at some $\Lambda$ is related to the average importance ratio over posterior samples for event $n$:

\begin{equation}
    \ev^n_{\text{Lag,Short}}\left(\Lambda\right)
    =
    \ev^n_{\text{Lag},\emptyset} 
    \times
    \text{E}_{n,\text{Posterior}}\left[
        \frac{\pi^n\left(\Delta t \vert \Lambda\right)}{\pi_\emptyset\left(\Delta t\right)}
    \right]
    .
    \label{eq: ZLambdashort_postexp}
\end{equation}
Here, $\ev^n_{\text{Lag,Short}}$ is the evidence for the prior-informed model up to the limit of $\tmax$, as our fiducial pre-fit samples do not extend beyond this bound. 

There is also the complication that our uniform prior fits extend up to a maximum range of $\Delta t_\mathrm{Max}=\tmax$, while the log-normal $R-L$ prior has no upper bound. For short lags like seen in the \hbeta sample this is not a major issue as very little probability is contained at large $\Delta t$, but, as examined in \paperone, the \qm{ceiling} on fit lags has a marked impact on the apparent slope of the $R-L$ relationship for \civ. Fortunately, we can get around this with a trick: a sufficiently long lag is indistinguishable from a complete lack of coupling, causing  the likelihood to flatten out at large $\Delta t$ and converge to the uncoupled likelihood i.e. $\mathcal{L}_\mathrm{Coup}(d^n\vert)\approx\mathcal{L}_\mathrm{Uncoup}(d^n\vert\Delta t)$ in the limit $\Delta t\rightarrow\infty$. Using this approximation (which we justify and examine the limits of in Appendix~\ref{app: num_stability_tmax}), we can estimate the full prior-informed evidence integral as:
\begin{equation}
    \ev^n_{\text{Lag}}=\ev^n_{\text{Lag,Short}} + \ev^n_{\text{NoLag}} \cdot \int(\pi^n\left(\Delta t \vert \Lambda\right))d\Delta t
    \label{eq: ZLambda_shortpluslong}
    ,
\end{equation}
For brevity we write $\int(\pi^n\left(\Delta t\vert\Lambda\right))d\Delta t=k(\Lambda)$ where $k(\Lambda)=\frac{1}{2} \left(1 + \frac{\text{erf}(\Delta t_\mathrm{Max}-\mu_{R-L}^n(\Lambda)))}{\sqrt{2}\sigma}\right)$ for $\mu_{R-L}^n(\Lambda)$ defined per Equation~\ref{eq: R-L_prior} is the amount of prior probability above $\Delta t_\mathrm{Max}=\tmax$.

Rather than estimate the average importance ratio in Equation~\ref{eq: ZLambdashort_postexp} as a sum over posterior samples (i.e. $\text{E}_{n,\text{Posterior}}\approx \frac{1}{I_n}\sum_{i_n}{ \left[ \frac{\pi\left(\Delta t_n^i \vert \Lambda\right)}{\pi_\emptyset\left(\Delta t_n^i\right)} \right]}$ where $I_n$ is the number of samples), we instead estimate it as a weighted finite sum over nested samples for numerical stability:
\begin{align}
    \label{eq: postexp_nessum}
    \text{E}_{n,\text{Posterior}}\left[\frac{\pi^n\left(\Delta t \vert \Lambda\right)}{\pi_\emptyset\left(\Delta t\right)}\right] 
    &\approx \sum_{i_n}{\frac{\Delta\mathcal{Z}_{n,\text{Lag},\emptyset}^i}{\mathcal{Z}_{n,\text{Lag},\emptyset}}\frac{\pi\left(\Delta t_ n^i \vert \Lambda\right)}{\pi_\emptyset\left(\Delta t_n^i\right)}}
    \nonumber
    \\ &=  \sum_{i_n}{w_i^n\frac{\pi\left(\Delta t_ n^i \vert \Lambda\right)}{\pi_\emptyset\left(\Delta t_n^i\right)}}
    ,
\end{align}
Here $w_i^n=\frac{\Delta\mathcal{Z}_{n,\text{Lag},\emptyset}^i}{\mathcal{Z}_{n,\text{Lag},\emptyset}}$ are the nested sampling to posterior weights for nested sample $i$ of event $n$. We demonstrate that this is a vastly more numerically stable approach at no additional cost in Appendix~\ref{app: num_nestsampstability}.

With all of these components, we can arrive at a simple finite summation expression for the marginal likelihood of a reverberating fraction $f$ and $R-L$ parameter set $\Lambda$ for a given line:
\begin{align}
    \label{eq: marglik_finitesum}
    \mathcal{L}\left(D\middle \vert \Lambda,f\right) 
    &\approx
    \prod_{n}{
        \mathcal{Z}_{n, \text{No Lag}}\sum_{i_n}{
            w_i^n\left[
            \begin{array}{l}
                \left(1-f\right) \\
                + 
                f \cdot \left({\rm \text{BF}}_{n,\emptyset}\frac{\pi\left(\Delta t_n^i \vert \Lambda\right)}{\pi_\emptyset\left(\Delta t_n^i\right)}+k(\Lambda)\right)
            \end{array}
            \right]
        }
    }
    \nonumber
    \\ &\propto 
    \prod_{n}{
        \sum_{i_n}{
            w_i^n\left[
                \begin{array}{l}
                     \left(1-f\right)  \\ 
                     +f \cdot \left( {\rm \text{BF}}_{n,\emptyset}\frac{\pi\left(\Delta t_n^i \vert \Lambda\right)}{\pi_\emptyset\left(\Delta t_n^i\right)} + k(\Lambda)\right)
                \end{array}
            \right]
        }
    }
    .
\end{align}

This formulation has a number of advantageous properties. Most notably, the numerical summation means that we have reduced the number of free parameters to only $4$ (the $3$ parameters in $\Lambda$ and the reverberating fraction $f$) and at least for our data this low dimensional distribution is smooth and unimodal. This allows us to fit the entire model with simple MCMC methods, in our work using a combination of the Sample Adaptive Sampler \citep{Zhu_2019_SAsampler} and Ensemble Slice Sampler \citep{Karamanis_2021_ESS} as implemented in \numpyro \citep{numpyro}. Secondly, the summations and products in Equation~\ref{eq: marglik_finitesum} are embarrassingly parallel, meaning this can be trivially distributed over parallel computing jobs. Thirdly, it is  easily differentiable, meaning that more complex gradient-based fitters like HMC or SVI are viable if the framework is extended to higher dimensional models. Finally, it requires only access to the nested samples of fiducial RM fits, meaning the complexity and computational cost is front loaded into the calculation of a data product we already have available from a more traditional RM analysis.

\subsection{Using Hierarchical Model Outputs to Constrain Single Sources}
\label{sec: methods_constrain_single}
Our calculation method for fitting our hierarchical model described in Section~\ref{sec: methods_practical_calc} marginalises out the individual source properties in favour of constraining the population parameters. Nonetheless, we can still use it to derive individual source properties, here presented with numerical recipes for use after population posterior fitting is complete. We present three measures of interest for each source, along with instructions on how to calculate them numerically: the posterior lag distributions, the Bayes factor for the lag (coupled vs uncoupled models), and the marginal probability of a source exhibiting a lag.

\paragraph{Posterior Distributions}
Once our $R-L$ parameters are constrained, we have a probabilistic set of $R-L$ informed priors with which we can re-weight the pre-fit uniform-prior lag samples that we used as inputs to our hierarchical model. These provide refined constraints on $\Delta t^n$ for all sources in our sample.

Because the uniform prior extends to an upper bound $\Delta t=\tmax$, this is a two part process: Where the prior is well below $\Delta t_\mathrm{Max}$ getting lag constraints are a simple matter of re-drawing the fiducial posterior samples weighted by $\pi(\Delta t^n\vert\Lambda) / \pi_\emptyset(\Delta t)$, but above this limit we draw directly from our prior, as the likelihood is assumed to be constant per our assumption in Equation~\ref{eq: ZLambda_shortpluslong}. Because we already have estimates of the relative posterior weight  above and below the $\tmax$, we can do weighted draws from these two sets of samples.
\begin{eqnarray}
    P(\Delta t^n \vert d^n,&&\Lambda)
    \propto
    \int\pi^n(\Delta t^n\vert \Lambda) \mathcal{L}(d^n\vert\Delta t)\\
    &&\sim
     \redraw
     \begin{cases}
        \redraw(\Delta t_{\mathrm{NS}}^n, \mathcal{R}(w^n_j\Delta t_{\mathrm{NS}})) , \; \BF{\Delta t<\Delta t_\text{Max}}^n(\Lambda) \\
        \pi(\Delta t^n\vert\Lambda) \cdot H(\Delta t-\Delta t_\text{Max}) , \; k^n(\Lambda)
    \end{cases} 
    \nonumber
    ,
    \label{eq: single_lagredraw}
\end{eqnarray}
where $H$ is the Heaviside step function. To marginalise over the uncertainty in the population parameters, we repeat this for a set of $\Lambda$s drawn from our posterior fits, weight by $P_\mathrm{Lag}$ (method described below) and make a final redraw from the collective sample. The result is a set of lags, $P(\Delta t^n \vert d^n)$, that is marginalised over all uncertainties in all non-lag parameters at the source and population level.

\paragraph{Bayes Factor \& Marginal Probability of Being Coupled}
In addition to constraining the lag, we can also calculate our confidence in the lag by finding the per-source Bayes factors under our new $R-L$ informed priors. The Bayes factor is of interest in that it gives us a goodness of fit for the coupled model for a given source independent on $f$. It also is needed for posterior marginalisation and uncertainty estimates (see Appendix~\ref{app: num_stability_RLpropagation}).
\begin{equation}
    \BF{Lag}(\Lambda)=\BF{\Delta t<\Delta t_\text{Max}}^n(\Lambda)+(1-k^n(\Lambda))
    .
\end{equation}
The Bayes factor is also useful for evaluating the performance of the model, i.e. how much our information we gain from our $R-L$ informed prior (see an example in Figure~\ref{fig: bayeschange}).

Given that $f$ also acts as a population informed prior on whether or not a source is coupled, we can also turn this into a posterior probability on the source containing a lag:
\begin{eqnarray}
    P^n_\mathrm{Lag}(\Lambda,f)
    &=&
    \frac{f\times( \BF{Lag}(\Lambda))}{f\times( \BF{Lag}(\Lambda))+(1-f)}
    \label{eq: plag}
\end{eqnarray}

Both $\BF{Lag}(\Lambda)$ and $P^n_\mathrm{Lag}(\Lambda,f)$ depend on population properties. Marginalising over the uncertainty in $\Lambda$ and $f$ is as simple as averaging over many posterior samples from the hierarchical fit. Where we present these values without an argument (e.g. $\BF{Lag}$ or  $P^n_\mathrm{Lag}$) we refer to these marginal values. For testing how these might change subject to tightening constraints on our population parameters, we can also evaluate the degree to which they change over these posterior samples. Where we write these with a superscript \qm{$\mathrm{Med}$}, it means the median over posterior samples, and where we use the superscripts \qm{$\mathrm{Hi}$} and \qm{$\mathrm{Lo}$} it we refer to the uncertainties relative to the median from the \pone and \ptwo percentiles.

\section{Results from this New Method When Applied to the \ozdes Sample}
\label{sec: results}

In this section, we demonstrate the performance of our new method by applying it to the \ozdes RM sample. By comparing to the more traditional cut-and-constrain approach of \paperthree, we can demonstrate its markedly improved performance. In Section~\ref{sec: results_RL_constraints} we show our updated constraints on the $R-L$ relation for all lines compared with constraints from past \ozdes analyses, and briefly discuss what the performance of different light curve models tells us about AGN variability and BLR response functions. In Section~\ref{sec: results_new_lags_and_masses} we present our new sample of lags and SMBH masses, and in Section~\ref{sec: results_reverb} we discuss our new constraints on the number of AGN that present clear reverberation lags.

\subsection{Constraints on the $R-L$ Relationship \& Comparison with Traditional Methods}
\label{sec: results_RL_constraints}
In this section we present our constraints on the $R-L$ relation for \hbeta, \mgii and \civ, with the final fits for our fiducial light curve model summarised in Table~\ref{tab: R-L_Constraints}. We also discuss how the ability of our other light curve descriptions to fit an $R-L$ relation, the simple DRW and the DRW + Jitter models, using these to examine differences in the characteristics of each line sample's variability. For a comparison of these new $R-L$ constraints with past \ozdes and \sdss fits, see Appendix~\ref{app: context} and Figure~\ref{fig: contours_compare_lit} in particular.

We also take this chance to demonstrate the efficacy of our method by contrasting it with the results acquired from cut-and-constrain methods using the \gold sample from \paperthree. When making these cut-and-constrain fits with MCMC, we use the $R-L$ model we outline in \paperone, including the accounting for a $1250 \dayu$ rest-frame lag ceiling for the \civ sample described in that work.

\begin{table*}
    \centering
    \begin{tabular}{c|c|c|c|c|c|c}
        Line &  \makecell{Slope\\$\alpha$} & \makecell{Offset\\$\beta$} & \makecell{$\alpha$/$\beta$\\Corr.} & \makecell{Scatter\\$\sigma$} &  \makecell{Reverb.\\Frac. $f$} & \makecell{$\sigma/f$\\Corr.} \\\hline
\hbeta & $0.63_{-0.20}^{+0.22}$ & $-0.59$ & $1.47_{-0.12}^{+0.11}$ & $0.48_{-0.08}^{+0.10}$ & $0.91_{-0.06}^{+0.04}$ & $0.27$\\\hline
\mgii & $0.43_{-0.35}^{+0.28}$ & $-0.82$ & $1.98_{-0.16}^{+0.16}$ & $0.28_{-0.07}^{+0.08}$ & $0.09_{-0.02}^{+0.02}$ & $0.17$\\\hline
\civ & $0.13_{-0.15}^{+0.15}$ & $-0.90$ & $1.83_{-0.14}^{+0.14}$ & $0.45_{-0.05}^{+0.05}$ & $0.74_{-0.06}^{+0.06}$ & $0.25$\\\hline
    \end{tabular}
    \caption{Summaries of our $R-L$ parameter constraints for \hbeta, \mgii and \civ based on applying our Hierarchical model to the \ozdes sample. Correlations assume Gaussian distributions for the parameters, and uncertainties are listed from the \pone and \ptwo percentiles relative to the posterior median. All fits here are for our fiducial light curve model, and without removing noisy sources as we discuss in Appendix~\ref{app: modelling_stationary}. The \hbeta relation is fit with a luminosity offset of $10^{44} \ \mathrm{erg/s}$, while the \mgii and \civ relations are fit for offsets of $10^{45} \ \mathrm{erg/s}$.}
    \label{tab: R-L_Constraints}
\end{table*}

\begin{figure}
    \centering
    \includegraphics[width=1.0\linewidth, trim={0.25cm, 0.5cm, 0.25cm, 0.0cm}, clip]{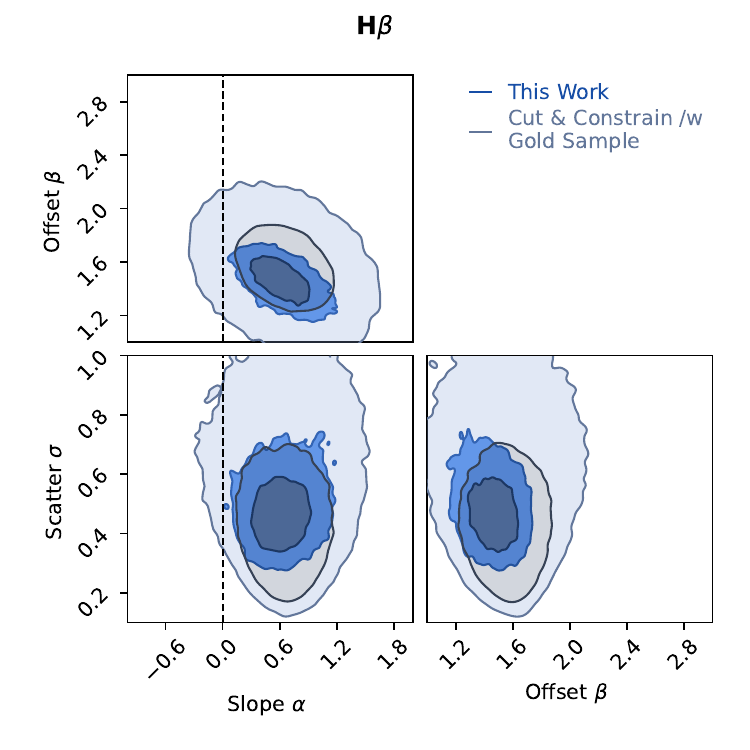}
    \caption{Posterior contours for our $R-L$ parameter constraints for \hbeta, compared to constraints using the cut-and-constrain approach using the results of \paperthree's \gold sample. The case of no slope, i.e. $\alpha=0$, is marked with a dotted line.}
    \label{fig: contours_vs_cnc_Hbeta}
\end{figure}

\paragraph{Constraints for \hbeta}
Compared to the cut-and-constrain method, \hbeta sees a significant improvement over traditional methods. Shown in Figure~\ref{fig: contours_vs_cnc_Hbeta}, we see a more than $\sim2\times$ improvement in $R-L$ constraints for this line, equivalent under $\sqrt{N}$ scaling to an $\approx4\times$ increase in source count. With this method, we are able to conclusively demonstrate the existence of the lag-luminosity scaling (i.e. constrain the $R-L$  slope $\alpha>0$ at $2\sigma$) using only the limited data of the \ozdes \hbeta sample. The old \ozdes \hbeta sample was unable to constrain a scaling at all \citep[see][and \paperone]{OzDES-Malik_2023}, and even with our improved source-by-source analysis in \paperthree the \ozdes data was insufficient to conclusively show this scaling. We show this clear slope in our $R-L$ fit, along with the marginal distributions for the rest-frame lags that it fits and is informed by, in Figure~\ref{fig: RL_violin}.

\begin{figure}
    \centering
    \includegraphics[width=1.0\linewidth]{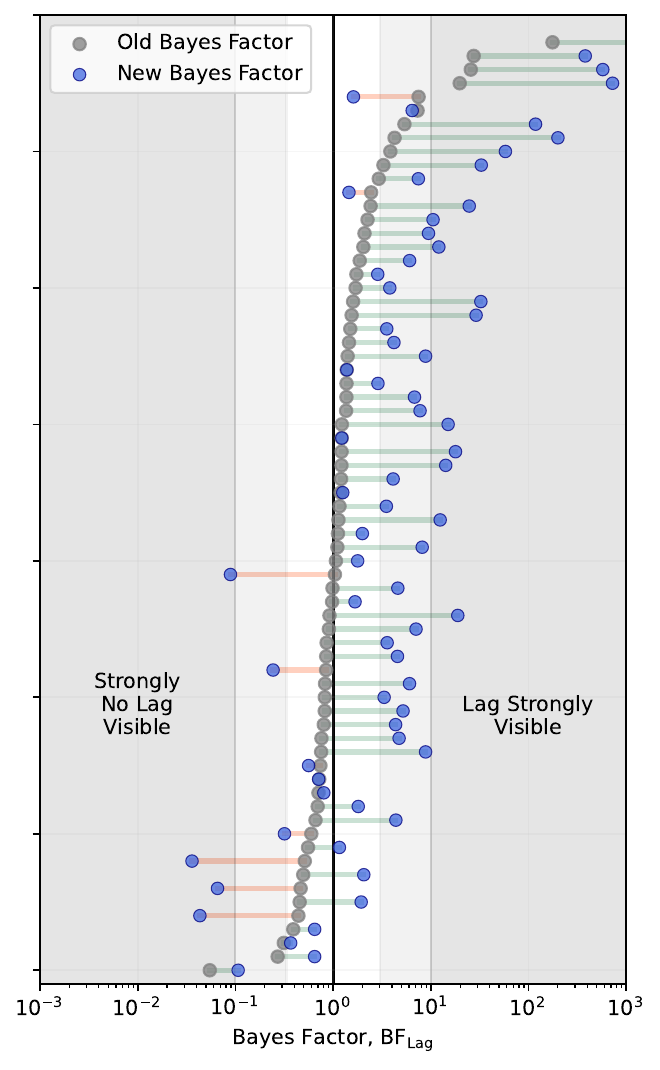}
    \caption{A demonstration of how our Bayes factors / confidence in lag recoveries changes by including an $R-L$ informed prior for the \hbeta sample under the simple DRW model, with shaded regions indicating typical bounds for Bayes factor confidence margins ($1:3$ and $1:10$ for \qm{interesting} and \qm{convincing} ratios). Approximately $3$ sources have their confidence weakened, but by and large the trend is an increase in resolution, i.e. a magnification of our confidence in whether a signal does or does not contain a lag.}
    \label{fig: bayeschange}
\end{figure}

\hbeta also sees a drastic improvement in the number of recovered lags. Finding an effective source count by summing $P_\mathrm{Lag}^n$ for the \hbeta sources, we have an effective lag recovery count of $\sim63/69$ sources, $46$ of which are extremely high quality ($P_\mathrm{Lag}^n \ge 95\%$). By contrast, cut-and-constrain approaches in \paperthree yielded information from only $12/77$ of the targets at the \gold confidence level discarding the rest. The reasons for this are twofold: firstly (discussed in Section~\ref{sec: results_reverb}) almost all of the \hbeta sources are expected to have clear lags, meaning that even in the limit of extremely weak lag constraints we still constrain lags with population informed priors in a single-epoch like way. Secondly, because the $R-L$ prior is log-normal and \hbeta is associated with the lowest lags (See Figure~\ref{fig: priordens_estimates_plot}), it is in this sample that these population priors give the greatest improvement in signal clarity. This is shown though the frame of changes in Bayes factor relative to uninformative priors in Figure~\ref{fig: bayeschange}.

\hbeta is the least sensitive of all three lines to the choice of light curve model: including jitter or response smoothing does not markedly change our $R-L$ relation fits, aside from the inclusion of tophat smoothing preferring a slightly higher scatter (higher $\sigma$) and slightly longer lags (higher $\beta$) compared to the other models. This indicates that either the \hbeta light curves do not display significant decoupled noise or smoothing, or that their short lags make them so easily recoverable that model misspecification has a negligible impact. 

\begin{figure}
    \centering
    \includegraphics[width=1.0\linewidth, trim={0.25cm, 0.5cm, 0.25cm, 0.0cm}, clip]{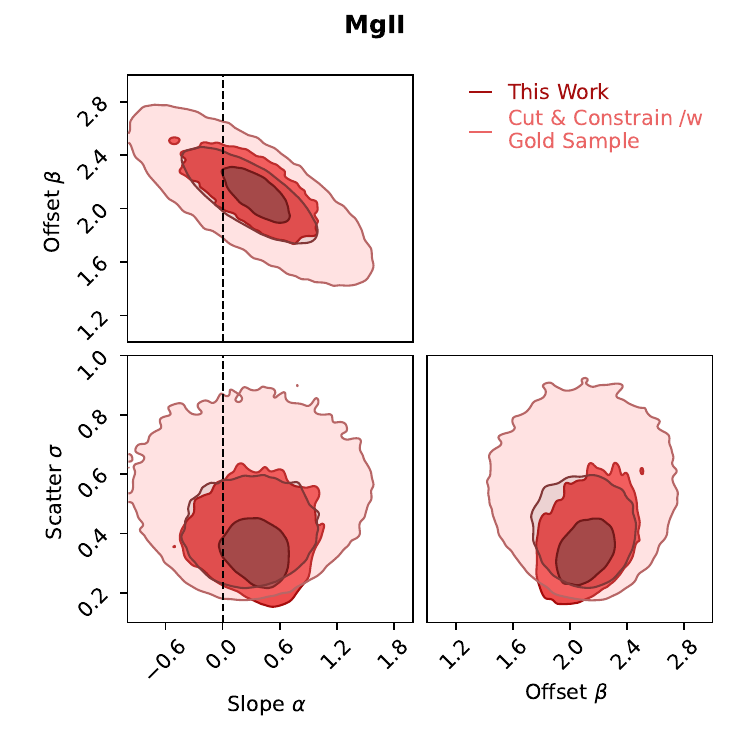}
    \caption{Posterior contours for our $R-L$ parameter constraints for \mgii, compared to constraints using the cut-and-constrain approach using the results of \paperthree's \gold sample. The case of no slope, i.e. $\alpha=0$, is marked with a dotted line.}
    \label{fig: contours_vs_cnc_MgII}
\end{figure}

\newpage
\paragraph{Constraints for \mgii}
Similar to \hbeta, the \mgii sample sees an $\sim2\times$ increase in $R-L$ constraining power as compared to cut-and-constrain techniques in \paperthree (see Figure~\ref{fig: contours_vs_cnc_MgII}), though with \ozdes alone the constraints are too broad to fully exclude an unscaled fit (i.e. $\alpha=0$, see Figure~\ref{fig: contours_vs_cnc_MgII}). The lack of resolution is due to a low effective source count owing to \mgii having only a very small reverberating fraction, coupled with the broadness of its $R-L$ relation meaning less prior information is supplied from the prior. Nevertheless, in this hierarchical model we go from $9$ \gold standard recoveries with traditional techniques to an effective lag measurement count of $38.4$, an increase of more than $4\times$.

The \mgii $R-L$ relation can be fit if we allow for smoothing in the continuum (our fiducial model) or by adding additional jitter noise, but cannot be fit when using a simple DRW alone. As noted in \paperthree, allowing jitter in the continuum has a second-order smoothing effect that increases $\tau$ by smoothing out rapid variations. This suggests that the jitter model performs well by allowing the \mgii variability timescales to be long, but not necessarily longer than the continuum, though we avoid any solid conclusions without further analysis. Fits with the jitter model are consistent with fitting for continuum smoothing but with markedly larger statistical uncertainty. 

We discuss this further in Section~\ref{sec: widerimpact_modelling_LCmodels} and Appendix~\ref{app: modelling_modelsensitivity}, but the \mgii sample seems to be the most sensitive to light curve model choice: it has, for some balance of data reduction or astrophysical reasons, the most difficulty being describe by a simple call and response model. With our sampling here, though our effective sample size increases by a factor of $\sim 3$, the number of extremely reliable recoveries decreases from $9$ in \paperthree (those with $\mathrm{FPR}\le5\%$ in a negative lag test) to only $3$ in this work ($P_\mathrm{Lag}^n\ge95\%$).

\begin{figure}
    \centering
    \includegraphics[width=1.0\linewidth, trim={0.25cm, 0.5cm, 0.25cm, 0.0cm}, clip]{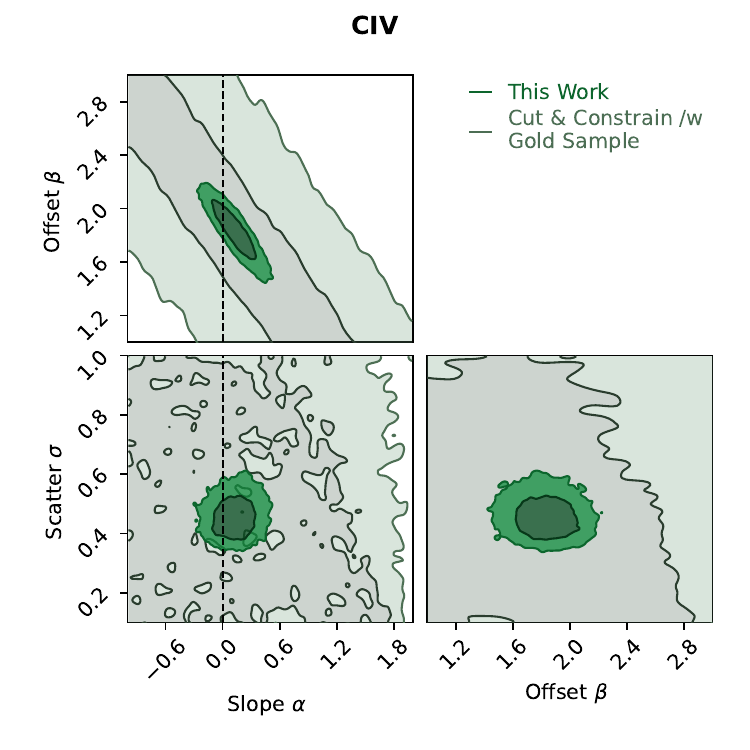}
    \caption{Posterior contours for our $R-L$ parameter constraints for \civ, compared to constraints using the cut-and-constrain approach using the results of \paperthree's \gold sample. The case of no slope, i.e. $\alpha=0$, is marked with a dotted line.}
    \label{fig: contours_vs_cnc_CIV}
\end{figure}

\paragraph{Constraints for \civ}
Unlike the \hbeta and \mgii samples, which markedly improve on what would at least be reasonable $R-L$ fits, applying our method to the \civ sample allows us to constrain $R-L$ parameters where previously the sample was too small to make any meaningful statement about the population. In Figure~\ref{fig: contours_vs_cnc_CIV} we show this improvement: with $3$ \gold recoveries in \paperthree's cut-and-constrain sample there was too little information to make any constraints on the $R-L$ parameters whatsoever. With our hierarchical model, we are able to put meaningful bounds on the population properties, though the narrow luminosity range and finite length lag ceiling of the \ozdes survey means that the slope and offset are too correlated to properly identify the $R-L$ scaling. As shown in \paperone, this degeneracy can be broken by combining \ozdes with either more high redshift data, or with a low or high luminosity anchor to the population \citep[e.g. the collections of][]{Kaspi_2000,Kaspi_2007}. Our results here cannot exclude $\alpha=0$, but does demonstrate a shallow slope inconsistent with a $\alpha=0.5$ scaling expected from naive photoionisation models.

In \paperthree we found that the long lags of the \civ sample made it very difficult to measure them reliably or in large quantities. With this hierarchical method, which also includes prior-informed single-epoch like information, \civ sees a nearly two order of magnitude improvement in lag recovery numbers, from only $3$ \gold standard recoveries to an effective sample size of $\sim208$ targets of the total $280$. This large yield of lags comes from our finding that $f_{\civ}\approx70-80\%$ (See Section~\ref{sec: results_reverb}), which means we can comfortably use the our $R-L$ relation for the majority of sources. Even restricting to the most reliable recoveries, those with $P_\mathrm{Lag}^n\ge95\%$, we still acquire $14$ extremely reliable lag recoveries.

The \civ sample is the most model-sensitive when fitting the $R-L$ relation; of our light curve models in Section~\ref{sec: RM_stats_LCmodel} a population fit can only be made when including smoothing in the response light curve. Unlike \mgii, which required only that the response have a long timescale, this indicates that the \civ light curves need to be fit at a distinctly different timescale to the continuum. This is in keeping with our findings from \paperone in which we found that $\sim35\%$ of \civ sources had either probable or clear smoothing in their response.

\subsection{New Sample of Lags \& SMBH Masses}
\label{sec: results_new_lags_and_masses}
Using the approach outlined in Section~\ref{sec: methods_constrain_single} we can constrain lags from all $796$ sources in our sample, as well as marginal probabilities of those sources having coupling between the continuum and response light curves for use to draw a lag from. Fitting for our fiducial light curve model, and limiting to only decent quality recoveries, i.e. those with more than a $50\%$ marginal probability of containing a lag per Equation~\ref{eq: plag}, we recover $64$ \hbeta lags, $26$ \mgii lags and $257$ \civ lags. The low source count for the \mgii is a result of its low reverberating fraction, which we discuss in Section~\ref{sec: results_reverb}. We list these higher quality recoveries in Appendix~\ref{app: final_sample}, and the full un-truncated list of lags for all reliabilities are available in the online data release (see Section~\ref{sec: data_release}).

\begin{figure*}
    \centering
    \includegraphics[width=1.0\linewidth]{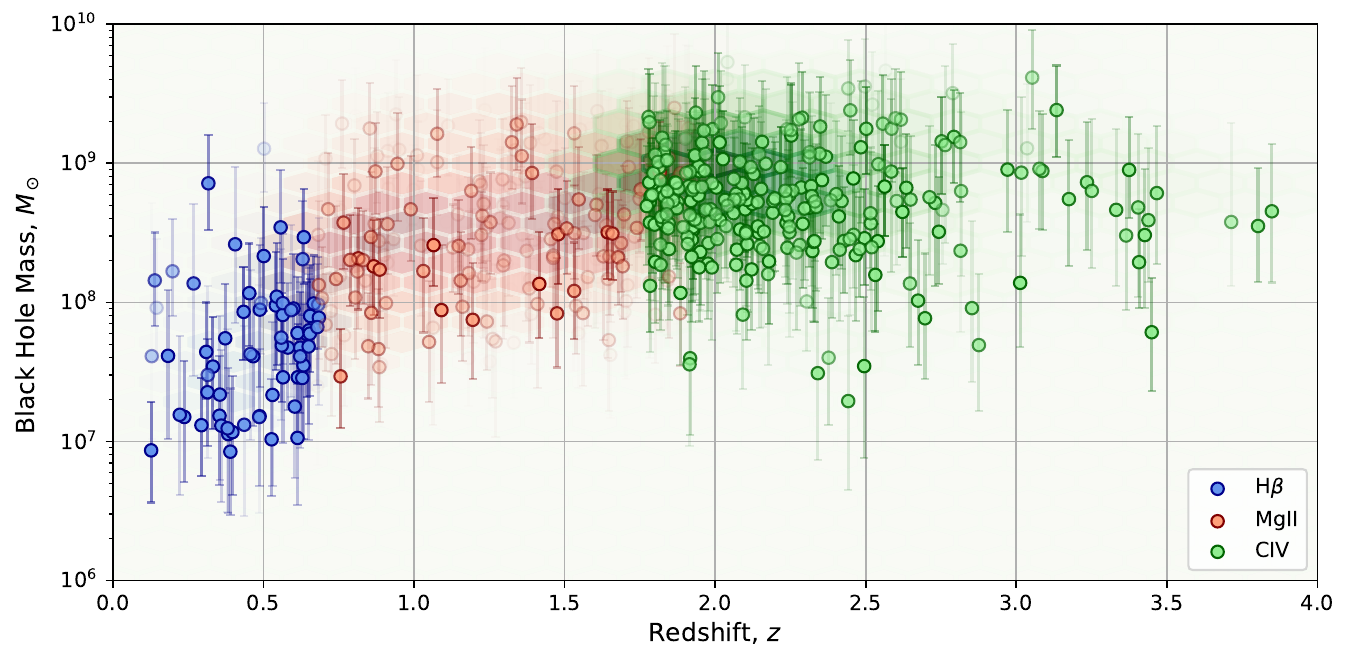}
    \caption{Our recovered masses over redshift, with samples coloured by line type and opacity scaled by $P_\mathrm{Lag}^n$, the marginal probability of containing a lag in Equation~\ref{eq: plag} Error bars represent \pone and \ptwo posterior percentiles after marginalising over uncertainties in $\Lambda$, virial factor, velocity line width and source lag. Underlaid with coloured shading are hexbins demonstrating the boot-strapped mass distributions for each line as a function of redshift.}
    \label{fig: new_masses}
\end{figure*}

From this sample we can also constrain $764$ SMBH masses from sources with well resolved velocity line widths, with $475$ at $\ge50\%$ marginal probability of a having a lag detection. For our mass estimates, we use the virial factor of $\log_{10}(f)=0.62 \pm 0.07 \pm 0.31$ from \citet{SDSS-Shen_2023}, the first uncertainty here being statistical error and the second being the scatter in the population. We show our new lag sample in Figure~\ref{fig: new_masses}, in which we can see the same downturn in masses towards low redshift that we observed with the \ozdes single epoch sample in \paperone, nicely demonstrating the effect cosmic downsizing in the AGN population \citep{Barger_2005_downsizing, Vestergaard_2009_downsizing, Kelly_2010_downsizing, Fanidakis_2011_downsizing}. Taking a weighted sum of our recoveries to get an effective number of mass recoveries, i.e. $\sum_nP^n_\mathrm{Lag}$ for all sources with a mass estimate, we arrive at an effective count of $297$ mass recoveries. Compared to the $24$ \gold standard masses recovered in \paperthree with traditional non-hierarchical methods, this is an order of magnitude increase in recoveries. This technique has also extended the range of our sample, with $132$ effective mass measurements at $z\ge2.0$, while the same \ozdes sample had only $2$ \gold standard recoveries in that same range. The most distant source for which we measure a mass is the \civ lag for \ozdes source $2940369952$ ($P^n_{\rm Coup}=0.83$). At a redshift of $z=3.849$, this is among the most distant sources for which a mass has been derived from RM, comparable to the most distant lag from \citet{SDSS-Shen_2023} at $z=4.29$.

\subsection{How many AGN Have Recoverable Lags?}
\label{sec: results_reverb}
In \paperthree we used the Bayes factors from the fiducial uniform priors to put estimate the fraction of reverberating AGN for each line, motivated by a low number of significant \mgii lag recoveries contrary to its large sample size. In that work, we noted that these were strict lower bounds, biased downwards by the uninformative nature of the uniform prior and its upper bound of $1500 \ \dayu$. Nonetheless, we found that, while the \hbeta and \civ could be shown to reverberate for most or all AGN in their sample, only a few percent of \mgii light curves encoded lags in the entire sample. In this work, we are equipped to examine the question of how many AGN reverberate in more detail. 

\begin{figure}
    \centering    
    \includegraphics[width=1.0\linewidth, trim={0.0cm 0.0cm 0.0cm 0.7cm}, clip]{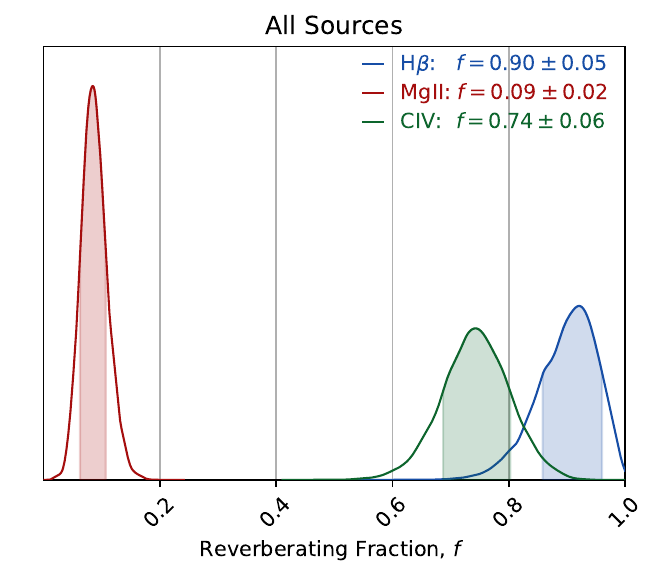}\\
    \caption{Posterior distributions for the underlying fraction of AGN, $f$, that exhibit lag-bearing coupled light curves between the continuum and response for all lines under our fiducial light curve model.}
    \label{fig: reverb_frac_constraints}
\end{figure}

Marginalising over our model hyper parameters while using our fiducial light curve model, we find good agreement with \paperone for \hbeta reverberating near to all of the time ($f_{\hbetamath}=0.90 \pm 0.5$), while \civ reverberates for $\sim \frac{3}{4}\%$ of our sample ($f_{\civ}=0.74\pm0.06$, see Figure~\ref{fig: reverb_frac_constraints}), an increase from \paperthree which found a lower estimate of about $\sim25\%$. These results are reasonably consistent among all descriptions of the light curve variability where we are able to fit an $R-L$ relation. The finding that $\sim100\%$ of \hbeta sources reverberate is comforting, as it means that past attempts at stacking or $R-L$ fitting for this line will be less biased by false positives. Using only our simple population modelling, $f_{\hbetamath} \approx 1.0$ is disfavoured, but we find in Appendix~\ref{app: modelling_stationary} that the small number of non-lag bearing \hbeta sources can be explained as having noisy white-noise response light curves. This suggests that the response decoupling of the \qm{BLR holiday} \citep{Goad_2016} do not complicate our \hbeta recoveries too severely.

Of greatest interest is the reverberating fraction for \mgii, which does see a marked increased from $\approx 3-5\%$ in \paperthree to $\approx9\%$ with more thorough modelling, but still shows only a fraction of sources encode resolvable lags. This low reverberating fraction is consistent between all models except the simple-DRW model, which has issues rejecting poor fit lags. In Appendix~\ref{app: modelling_modelsensitivity}, we show that the \mgii sample shows the most signs of having some variability in its continuum or response that cannot be explained with a simple call and response model, and this low count may indicate that the reverberating component of the response is subdominant in a large number of these sources. The fraction of $f\approx9\%$ is consistent even when allowing for extra calibration error with jitter, suggesting that, while data calibration might play a part in this paucity of recoverable lags, it is unlikely to be the dominant factor. 

We subdivide the non-reverberating AGN into those with structured and unstructured responses in Appendix~\ref{app: modelling_stationary}, and find that $\approx 2/3$ of the \ozdes \mgii sources have no meaningfully structured variability. This is contrary to the \hbeta and \civ samples which are consistent with clear signal in $100\%$ of targets. To within error, only $1/3$ of \mgii sources have non-stationary response light curves, and of those only $1/3$ contain lags. This is consistent with the finding of \paperthree, in which we found that removing sources of low response SNR (i.e. low $\BF{Struc,Resp}$ in Section~\ref{sec: RM_stats_BF}) achieved a maximum reverberation fraction of $f_{\mgii}\approx28\%$. This, along with the robustness of the finding to light curve model, suggests that this is an unavoidable feature of the \ozdes RM sample, and potentially of the \mgii AGN population at large.

\section{Impact of This Method on the Wider Field}
\label{sec: widerimpact}

In Section~\ref{sec: results} we demonstrated how our method significantly improves on lag measurements and population constraints for the \ozdes BLR RM sample, but it is worth putting this work in context with the broader field of AGN RM. In Section~\ref{sec: widerimpact_past}, we briefly discuss how our new $R-L$ fits compare to past \ozdes fits and the similar industrial-scale, high-redshift results of \sdss. In Section~\ref{sec: widerimpact_future}, we examine how this work can be applied to ongoing or upcoming surveys, including RM studies in regions other than the BLR. In Section~\ref{sec: widerimpact_modelling_LCmodels} and Section~\ref{sec: widerimpact_modelling_popmodels} we discuss how our method can be adapted to more expressive models of AGN variability and their population, and suggest some potential areas of future research.

\subsection{Comparison with Past Work}
\label{sec: widerimpact_past}
To better contextualise how our constraints fit into the broader endeavour of RM-based $R-L$ fitting, we also compare to $R-L$ fits for previously published \ozdes lags (i.e. those from \citet{OzDES-Malik_2023}, \citet{OzDES-Yu_2023} and \citet{OzDES-Penton_2025}), as well as the combined-survey $R-L$ fits from \paperone (based in part on those same past \ozdes analyses). We show the full comparative corner-plot contours for these fits in Appendix~\ref{app: context}, but summarise the key points here.

We generally see agreement when comparing to the original \ozdes analysis (though with worse constraints for \mgii primarily due to the much lower number of effective lags from the low reverberating fraction discussed in Section~\ref{sec: results_reverb}). The constraints for \hbeta are within statistical uncertainty compared to \citet{OzDES-Malik_2023}'s analysis of the same data, but identifying an $R-L$ scaling much more strongly and preferring overall higher intrinsic scatter in the population (higher $\sigma$). For \mgii we see similar agreement with \citet{OzDES-Yu_2023}'s analysis, but with markedly wider uncertainties and a preference for higher intrinsic scatter. For \civ, our results show a clearly higher intrinsic scatter than recovered by \citet{OzDES-Penton_2025} and \citet{OzDES-Hoormann_2019}'s \ozdes analysis, but recover almost identical constraints on the slope and offset. For all three lines, the statistical uncertainty is high enough that our new results are not in tension with the multi-survey combined $R-L$ fits that we presented in \paperone.

\begin{figure}
    \centering
    \includegraphics[width=1.0\linewidth]{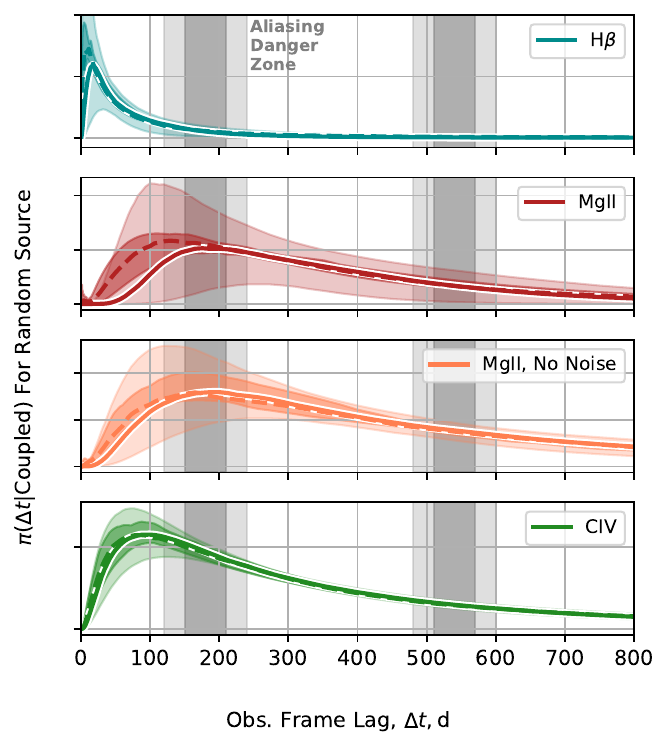}
    \caption{Our estimates of the rest-frame lag distributions based on our $R-L$ prior constraints, marginalised over the luminosity and redshifts of the \ozdes sample for each line and with grey shaded regions showing $\pm30\dayu$ and $\pm60\dayu$ intervals either side of the $180 \dayu$ and $540 \dayu$ aliasing danger zones. The dark shaded regions are the \pone and \ptwo uncertainty bounds varying only the population parameters, the light shaded regions are the same but while also bootstrapping over the sample. The solid curves are the median probabilities after varying population properties and bootstrapping, and the dotted lines are the same without bootstrapping. The red \mgii panel shows the results if we fit the population in the normal way and the orange \qm{No Noise} panel is the same result if we remove low variability or stationary sources as described in Appendix~\ref{app: modelling_stationary}. We can see that both fits for the \mgii sample peak at $\approx180\dayu$.}
    \label{fig: priordens_estimates_plot}
\end{figure}

In \paperone, we noted that the past \ozdes RM lags showed significantly lower scatter than the results off \sdss in \citet{SDSS-Shen_2023}. In that work, we also showed that sub-sampling the \sdss data set to emulate \ozdes's old cut-and-constrain approach reduced this scatter to be consistent between the two. Across all three lines, our rigorous modelling here prefers higher scatter than previous \ozdes work, bringing our fits closer to the findings of \sdss for \hbeta and \mgii. For \civ, however, the scatter shown in the \sdss results is still drastically higher than anything shown by \ozdes. In \paperone, we suggested that the \sdss data release may have a higher than estimated false positive rate with their quality cuts, and in that work we showed that this tension disappeared if we fit their data with the same cuts as past \ozdes work.

While our results here do not contradict the $R-L$ constraints of any past \mgii work, our finding that only $f_\mgii < 10\%$ of \mgii sources have lags, and the low lag recovery rate this results in, does put our findings in conflict with other comparable \mgii studies. In Figure~\ref{fig: priordens_estimates_plot} we show one possible reason for the high recovery rates: with our population fits, we find that an \ozdes-like optical RM survey results in the most probable lag from \mgii sources being near the $180\dayu$ aliasing peak. This means that true lag recoveries very often look very similar to spurious ones. As we discuss in Appendix~\ref{app: modelling_modelsensitivity}, there is good evidence that the \mgii response light curves are in some way more complex than can be described by the simple call and response models that so easily fit for \hbeta and \civ lags, and this is likely an element in explaining this result. We suggest that a similar rigorous Bayesian hierarchical analysis be applied to the \sdss data set to examine how much this finding changes when examining a similar sample with a different data reduction pipeline.

\subsection{Applications to Other Surveys \& Science Goals}
\label{sec: widerimpact_future}

Though \ozdes is an industrial-scale RM survey, it is only one of many. It is of interest to examine how our pipeline here may be applied to other larger samples, like that of \sdss, or upcoming wide-field surveys like \desi or \tides. Our method may be equally applied to surveys of non-BLR RM such as the large-scale accretion disk studies made possible by surveys like \lsst \citep{Kovacevic_2022_LSSTRM} or ATLAS \citep{steyn_2026_continuumreverberationbrightquasars}

As we show in \paperone, analysis methodology is the dominant issue in the accuracy and reliability of $R-L$ fitting in industrial-scale RM, and past non-Bayesian means of gauging these reliabilities fail to agree with one another. Even when using Bayes factor significance measures in \paperthree, we found that traditional non hierarchical methods are insufficient in particular for the \mgii sample. We demonstrate why this is in Figure~\ref{fig: priordens_estimates_plot}: for an \ozdes-like optical survey the bulk of optically visible \mgii sources have lags that fall within seasonal gaps, meaning they are difficult to distinguish from random chance. This unavoidably lowers our ability to confidently resolve lags without hierarchical methods. Even for the \hbeta sample, which are the least impacted by the complexities of our time-domain window function, this method still yields a multiple times improvement in population fitting.

Our method's complete modelling of the window function means we can push past the lag ceiling of finite survey length. Though only \civ is markedly affected by this finite-survey length lag ceiling for the $6$-year \ozdes campaign, it will remain an important effect for all lines for \desi and \tides in their earlier years. In Appendix~\ref{app: context_forecasting} we estimate that non-hierarchical methods will be impacted by a lag ceiling for the \hbeta sample until these surveys reach a length of $\sim2-4$ years, and the \mgii sample for $\sim5-8$ years. As we show in \paperone, failing to model in these finite survey length effects can severely distort our $R-L$ fitting. 

In Appendix~\ref{app: num_futurescalability}, we demonstrate that our method, without alteration, should be applicable to surveys up to $100\times$ the size of \ozdes before numerical precision becomes a dominant issue. In that appendix, we also suggest that only a few simple alterations are needed to future-proof our method going beyond even this limit. The only area we suggest further work is needed is in better characterising the light curves of the AGN, which we discuss further in Section~\ref{sec: widerimpact_modelling}. 

Though the hierarchical model may be over-engineered for smaller surveys, we do suggest at least using a coupled / uncoupled mixture model to marginalise over outliers when fitting any $R-L$ parameters. As shown in \paperone and here, the lag ceiling and low reverb fraction in \civ and \mgii have major impacts and need to be properly modelled. As such, we suggest that a hierarchical model, or at least a rough accounting for finite survey length as we showed in \paperone, is a necessary component of any analysis of the \civ line or for short surveys. For indications of how long a survey must be before exceeding the lag ceiling for a particular line, see Appendix~\ref{app: context_forecasting}.

\subsection{Extensions to Modelling}
\label{sec: widerimpact_modelling}
Our method here applies to only a very simple version of RM: using call and response light curve models to fit a power-law $R-L$ relation. In this section, we briefly examine how our framework may be improved by using it in concert with more complex modelling of either the individual sources or of the population at large.

\subsubsection{Extensions to Light Curve Modelling}
\label{sec: widerimpact_modelling_LCmodels}

Though we test a variety of simple light curve models in this work (see Section~\ref{sec: RM_stats_LCmodel}) these are all only simple first-order corrections of the simple DRW model. For the \mgii and \civ samples, where the signal clarity of the lags is often poor, the limitations of our approximate models will have an impact on our lag recovery rates. To summarise a more thorough analysis in Appendix~\ref{app: modelling_modelsensitivity}, it appears that the \mgii light curves in particular have some level of additional noise, either from \feii contamination \citep[see][for \ozdes's calibration of these lines]{OzDES-Yu_2021} or from true variability in the emission line. Taken to its extreme, this would explain why the \mgii sample has so few structured light curves, and if this were a non-stationary noise process it would also explain why so many of the structured AGN show clear decoupling. A similar, if less severe, statement can be made for \civ and its $\sim25\%$ decoupled sources. The \hbeta sample and its clear lags do not seem to be as sensitive to our light curve model, but as we found in \paperthree the short lags it is associated with are most likely to be impacted by un-modeled short-timescale variability in the continuum. It is worth noting that these limitations are present in many RM techniques, and that our analysis here only makes such assumptions explicit.

We then repeat our suggestion in \paperthree that, now equipped with a statistical and numerical approach for approaching the problem rigorously, the next frontier in industrial-scale RM is in more thoroughly characterising the continuum and response variability in our targets. An area of particular interest is in modelling the response noise process ($n_r(t)$ in Equation~\ref{eq: RM_signalnoise}) as having some time-varying structure instead of as a uncorrelated white-noise. If decoupling in the \mgii sample is a true astrophysical process, we would expect it to have some timescale of variability associated with it. Failing that, we should at least quantify, through jitter-like noise models and more restrictive priors, the degree to which simple call and response models fail to describe the response light curves. We discuss in Appendix~\ref{app: num_futurescalability_persource} that our approach here should remain scalable for these more expressive light curve models, provided we fit only a single delay timescale.

\subsubsection{Extensions to the $R-L$ Model}
\label{sec: widerimpact_modelling_popmodels}
Our population model in this work is the simplest it can be: a simple log-linear $R-L$ relation. We stress that our hierarchical framework can be extended to any model of the AGN population, provided this model can be expressed as a closed form prior for the parameters, lag or otherwise, for the individual sources. As a simple example, there is good evidence that accretion rate is an additional predictor of \hbeta lags in low redshift AGN \citep{HBETA_Du_2016, HBETA_Hu_2021, HBETA_Hu_2025}. By labelling sources with their velocity line widths, we can fit their masses and accretion rates in our generative model, and then construct the population informed priors on lag in a way that includes these as inputs. Such population-level modelling may even extend as far as the fitting of cosmological parameters, as has been done as a post-processing step using existing lags in past works \citep{Cao_2022_cosmology}.

Though we are concerned mainly with tuning the single epoch predictors of lag in this work, this hierarchical model can easily be extended to fitting population models of non-lag parameters as well. For example, we know that the variability timescale of the AGN population varies with redshift \citep{Lewis_2023}, and there is nothing that stops us from refining our constraints on $\tau$ in the same way we refine our measurements on $\Delta t$ with the $R-L$ relation. In this work we normalise our light curves to arbitrary units, but future works might also consider modelling the flux variability of emission lines into their populations to separate our noisy and low variability stationary response curves. Similarly, if we infer mass and accretion rate as a part of our source-by-source modelling, our framework here could be used to constrain models of the SMBH mass population. With the \ozdes sample alone and our current light curve model limitations, we don't have the resolution to examine multi-axis predictors, but with more data an interesting next step would be to examine if the accretion-rate based break seen in the $R-L$ by the SEAMBH collaboration occurs in other lines or at high redshift \citep{Du_2015}. We discuss the numerical considerations of fitting such more expressive models in Appendix~\ref{app: num_futurescalability_popwide}.

\section{Conclusion \& Future Work}
\label{sec: conclusion}

We have demonstrated a new technique for performing reverberation mapping at scale which drastically outperforms traditional methods, using a fully Bayesian framework to extract previously discarded information. On existing data, we have demonstrated that this method more than doubles the constraining power of population-level models like the  radius luminosity scaling relationship, and improves the number of recovered RM lags and SMBH masses by more than an order of magnitude. We have shown that this method is numerically robust, and that it is comfortably scalable to much larger industrial-scale RM surveys. This method also presents a complete accounting for and handling of the aliasing problem in industrial-scale RM, as well as the issue of random chance false positives and the windowing function from the upper bound on recoverable lags for a finite length survey.

Using this method, we have confirmed our finding in \paperthree that, while nearly all \hbeta sources demonstrate clear RM lags, as does most of the \civ sample, only a fraction of the \mgii show evidence of reverberations. We identify that this can be broken down into two thirds of the sources in the \ozdes \mgii sample having no significant structured variability in their light curves at all, and two thirds of those that do still having no clear reverberation. Whether this complexity is a feature of the true \mgii variability or a question of data reduction and the complications of iron contamination is currently unclear.

We find evidence that typical modelling of RM light curves, with the response light curve being a shifted scaled and smoothed copy of the continuum, is a limiting approximation in large-scale RM analysis. On combination with the finding of \paperthree that there are short timescale noise processes in the DES continuum light curves not captured by the DRW model, we suggest that a valuable topic of future study is to more expressively characterise and model both the continuum and response light curves. We suggest that this would be of particular interest in the \mgii sample, in which the simple call and response models of current RM light curve fitting appear to be most misaligned with the true AGN variability.

We suggest that this hierarchical method be applied to other industrial-scale RM surveys to check for consistency in results, namely to the \sdss sample, identifying tensions in the same way as was done in \paperone. Where tensions exist in $\sigma$ and $f$, we might consider simultaneous fits for the $R-L$ slope and offset while allowing these noise / scatter parameters to vary and absorb differences in the data reduction pipelines and target populations. Rather than re-fitting for different permutations and combinations as we did in \paperone, we suggest fitting one survey at a time, using a Kernel Density Estimator (KDE) over the population parameters to estimate the likelihood shape, and then combine surveys with that evaluable surrogate likelihood.

\section*{Contribution Statement}
Analysis, programming, calculations: HM; Writing: HM; Original figures: HM; Editing: TMD, CL, PM, ZY, GFL, BP; \ozdes Project conception and coordination: TMD, CL, PM; Data generation and/or curation: all authors.

\section*{Acknowledgements}
We acknowledge and pay respect to the traditional owners of the land on which the University of Queensland, The University of Sydney, and Macquarie University are situated, upon whose unceded, sovereign, ancestral lands we work. We pay respects to their Ancestors and descendants, who continue cultural and spiritual connections to Country.

HGM, TMD, AP acknowledge support for early stages of this project from an Australian Research Council (ARC) Laureate Fellowship  (FL180100168) and for later stages from the ARC Centre of Excellence for Gravitational Wave Discovery, OzGrav (CE230100016). HGM has been supported by the Australian Government Research Training Program (RTP) award. We are grateful to the Australian public for enabling this science. 

Based in part on data acquired at the Anglo-Australian Telescope, under program Ab/2013B/012]. We acknowledge the traditional owners of the land on which the AAT stands, the Gamilaroi people, and pay our respects to elders past and present.

Calculations were made using \python \citep{VanRossum_2009_python} and with the aid of \texttt{numpy} \citep{harris_2020_numpy}. Plots and figures were generated with the aid of \texttt{matplotlib} \citep{Hunter_2007_matplotlib} and \texttt{chainconsumer} \citep{Hinton_2016_chainconsumer}. All initial lag fitting was done with \litmus, which is built using the probabilistic programming language \numpyro \citep{numpyro} and the \python framework \jax \citep{jax}. Fits are performed with the nested sampling package \jaxns \citep{Albert_2020_JAXNS}.

\section*{Data Availability}
\label{sec: data_release}
The results here are fit using data products from previous \ozdes RM papers. The emission line light curves are drawn from \paperone, and the outlier-cleaned response light curves are from \paperthree. We direct the reader to these papers for details on data availability.

As a part of this release, we include posterior samples for our $R-L$ constraints: \red{[ZENODO LINK GOES HERE]}

\bibliographystyle{apj}
\bibliography{bib_main, bib_OzDES, bib_LITMUS, bib_ozdeslitmus, bib_RL, bib_Hbetasources, bib_MgIIsources, bib_CIVsources}



\appendix
\counterwithin{figure}{section}


\section{Numerical Stability \& Accuracy}
\label{app: num}
In this appendix we provide the details on how we have validated the numerical stability and scalability of our method, with the conclusions mentioned in the text being based on the working here. In Section~\ref{app: num_nestsampstability}, we discuss why using raw nested samples instead of posterior samples is an important element in ensuring the accuracy and robustness of this hierarchical modelling. In Section~\ref{app: num_stability_ESS} we describe how we gauge the reliability of importance resampling on a per-source basis, and in Section~\ref{app: num_stability_RLpropagation} we show how these can be used to gauge the overall reliability of our population fits.

\subsection{Numerical Stability \& the Use of Nested Samples}
\label{app: num_nestsampstability}
Our calculation method relies on the accuracy of the Monte Carlo integral in Equation~\ref{eq: postexp_nessum}, a summation that can diverge when the denominator (sampling density) becomes small compared to the numerator (target density). In intuitive terms, changing the prior shifts the regions of high posterior weight, and we need to have a thorough enough sampling of this new region. If we do not do this, we can get posterior distributions that are poorly sampled, and our Monte Carlo integrals become either inaccurate or completely unstable. To avoid this, we need a sampling distribution that offers support to our new posterior distribution over many regions of hyper-parameter space. 

This sampling distribution has two requirements: it needs to be broader / have wider tails than the posterior so that the denominator in Equation~\ref{eq: postexp_nessum} does not become small compared to the numerator, but it also needs to be reasonably highly concentrated about the posterior peaks to adequately describe  its shape. Posterior samples from the fiducial prior maximally meet the second of these criteria, but fail in the first. By contrast, The nested samples from a nested sampling run (described in Section~\ref{sec: RM_stats_stats_nestedsampling}) meet both of these criteria. Because nested sampling starts by covering the entirety of parameter space and draws inwards across the run, saving samples all the while, we get some non-zero sampling of the entire domain. Because the NS run converges towards the fiducial posterior peaks, it also has high sample density at the regions of most interest.

If we fit our source-by-source non-hierarchical lag posteriors with nested sampling, then we already have access to a bank of nested samples for each source \qm{free of charge}. Using nested samples directly for importance sampling is strictly more stable and less lossy than using NS$\rightarrow$posterior redrawn samples. 

\begin{figure*}
    \centering
    \includegraphics[width=1.0\linewidth]{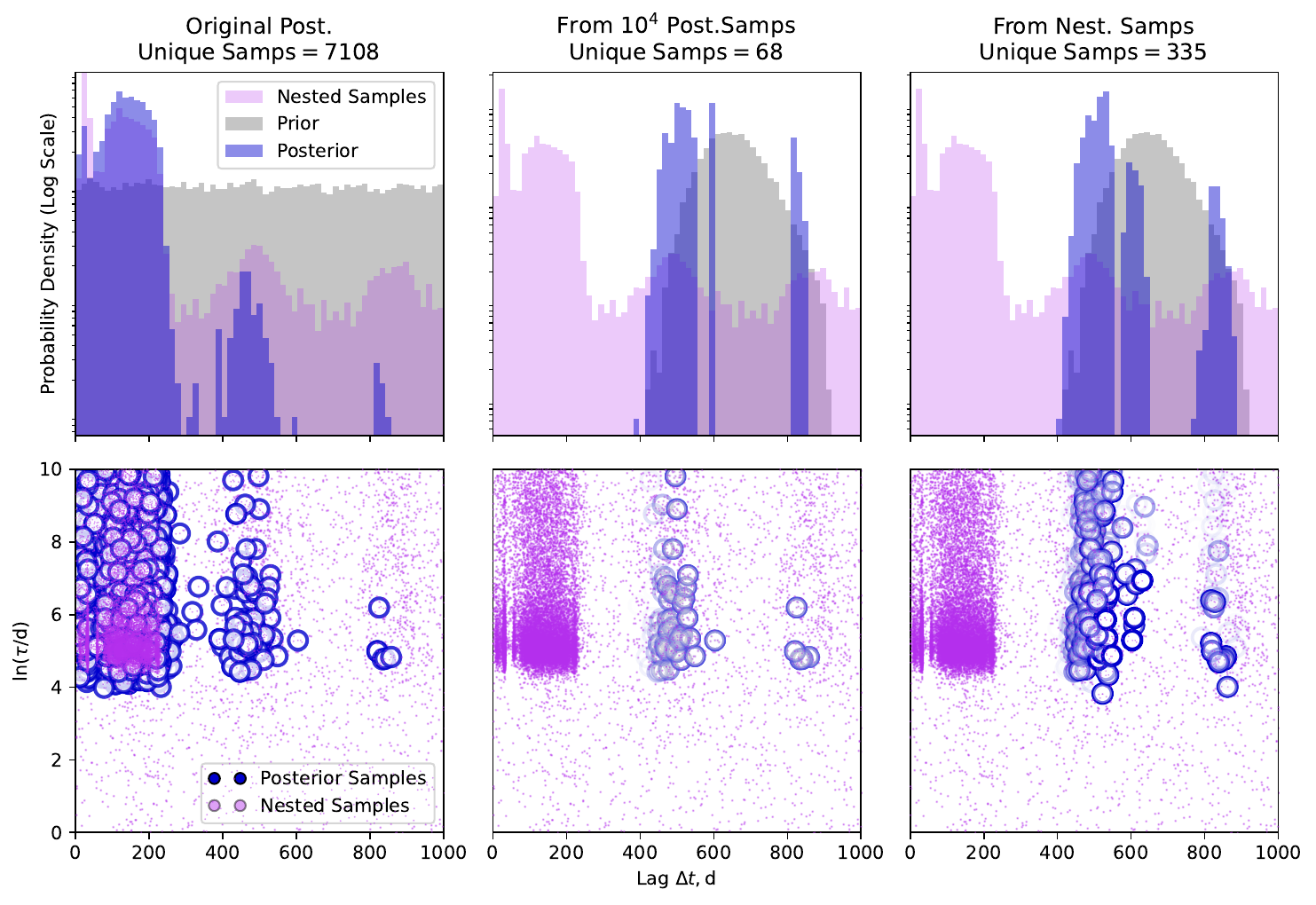}
    \caption{An example of how importance sampling from nested samples yields a better defined posterior than the posterior samples. The left panels show the posterior (navy), prior (grey) and nested sample (purple) distributions for the \hbeta lag of \ozdes source $2925351861$ as fit with the fiducial uniform prior, with the top row showing the marginal distributions for lag and the bottom row showing the a scatter plot of samples for the lag and log-timescale of variability. The posterior distribution here is after drawing $10^4$ posterior samples (navy) from the nested sampling run (purple). The middle panels shows the same results if we directly re-weight the posterior samples to an $R-L$ like prior with $\mu=\log_{10}{650 \dayu}$ and scatter of $\sigma=0.05 \; \mathrm{dex}$. Because the original posterior has thin sampling beyond $\Delta t\approx25 \dayu$, we have a sparse and incomplete picture of the re-weighted posterior, with $<100$ unique samples. The right hand panel shows the same fits if we instead importance sample directly from the original nested samples: our effective sample size has increased by a factor of roughly $\times5$.}
    \label{fig: reweight_example}
\end{figure*}

We give an example of why in Figure~\ref{fig: reweight_example}; when changing the prior drastically shifts the resultant posterior to a region of previously low posterior density, the sparsity of posterior samples give us a poor exploration of the new distribution, while the more thoroughly distributed nested samples give us markedly better information. This example is a deliberately extreme exaggerated edge case, and this stability is not a crucial issue for small sample sizes as, for most sources, the $R-L$ prior only narrows the posterior rather than dramatically shifting it. However, as our RM sample moves to the hundreds and thousands of sources, the odds of at least one source being numerically unstable becomes high. This is important when considering future scalability of our method for larger sample sizes.

We quantify this improved stability in Appendix~\ref{app: num_stability_ESS} with measures of \qm{effective sample size}, and also how we can quantify the numerical precision of our post-hierarchical lag constraints and model evidence integrals from these measures. We also estimate how numerical uncertainty propagates through to error in our $R-L$ constraints, and find that the numerical uncertainty from this method sits at $<1\%$ of the statistical error in the worst case. We discuss in some detail what considerations we need to keep in mind when scaling this method for large surveys in Section~\ref{sec: widerimpact_future}, but speaking broadly our approach of reprocessing nested samples allows our method to be comfortably applicable to samples up to $10\times$ the size of \ozdes without numerical concerns.

\subsection{Numerical Accuracy of Importance Sampling on a Per-Source Basis}
\label{app: num_stability}
In this Appendix section we discuss how we know our method is numerically accurate, both on a per-source basis in Appendix~\ref{app: num_stability_ESS} and how numerical uncertainty propagates through to our $R-L$ parameter fits in Appendix~\ref{app: num_stability_RLpropagation}.

\paragraph{Effective Sample Size}
\label{app: num_stability_ESS}
Here we describe some of the diagnostics we use for checking the numerical precision and stability of our importance sampling fits in the form of effective sample sizes, as well as some intuitive interpretations of what they mean.

For brevity we can write the importance sampling estimated evidence as the null model evidence, $\ev_{\fidp}$, multiplied by a sum over some set of $I$ sample weights $\{x_i\}$, i.e. $\ev_\text{Import.} = \ev_{\emptyset} \sum_i{x_i}$. We suggest two ways to estimate the effective sample size (ESS) of the importance sampled distributions. The first, which we call $\ESSo$, is a weighted sum of the number of samples:
\begin{equation}
    \ESSo=\frac{\sum_i{x_i}}{\max_i(x_i)}
    .
    \label{eq: ESS1}
\end{equation}
This measure tells us how many unique posterior samples we can expect to draw from the re-weighted distribution. If we attempt to draw $1000$ posterior samples when $\ESSo=100$, we would expect to redraw the same samples $10's$ of times over. This gives us a metric for well characterised our new lag-posterior distributions are. Our second measure, $\ESSt$, which tells us about how precise our estimates of the evidence / Bayes factors are, is defined:
\begin{equation}
    \ESSt=\frac{\sum_i{x_i^2}}{\left(\sum_i{x_i}\right)^2}
    .
    \label{eq: ESS2}
\end{equation}
If we make the coarse approximation of the terms in this sum being independent, we can estimate a worst case variance for the integrals, which can then be written as a relative error in terms of $\ESSt$:
\begin{align}
    \text{Var}(\ev_{\text{Import.}}) &\approx \ev_{\emptyset}^2
    ( \braket{x_i^2}  - \braket{x_i}^2) \nonumber\\
    & = \ev_{\emptyset} ^2
    \frac{(\sum_i{x_i}^2)}{I}( \sum{x_i^2}  - \sum{x_i}^2)  \nonumber\\
    & =  \ev_\text{Import.}^2\sqrt{\frac{1}{\ESSt} - \frac{1}{I}}\nonumber \\
    \frac{\text{Err}(\ev_\text{Import.})}{\ev_\text{Import.}} & \approx \frac{1}{\sqrt{\ESSt}}
    ,
\end{align}
where in the last line we have assumed $\ESSt \ll I$. This result matches our intuition for simple \qm{one on root $N$} scaling and gives a good yardstick for estimating the numerical precision of our importance sampled integrals. If we want the numerical error in a Bayes factor to be, say, less than $10\%$, then we need $\ESSt$ to be at least \qm{a few hundred}. We discuss how numerical uncertainty in these integrals propagates through to our $R-L$ parameter constraints in Section~\ref{app: num_stability_RLpropagation}. A more thorough treatment of this for small sample sizes can be found in \citet{heinzel_2025nottrustmontecarlo}

Because nested sampling runs follow the same exponential volume draw-in irrespective of their number of live points (above the threshold of reasonable convergence), increasing $\text{ESS}_{1/2}$ is as simple as re-running the nested sampling fit for a particular source in \litmus with an $N_\text{live}$ increased by the required factor. e.g. if the original run is performed with $N_\text{live}=1400$ and gives an average $\ESSt=50$, we re-run with $10\times$ as many live points to get $\ESSt\approx500$.

\subsubsection{Numerical Accuracy of $R-L$ Parameters}
\label{app: num_stability_RLpropagation}
While the effective sample size estimates in Section~\ref{app: num_stability_ESS} tell us about the numerical precision of our source-by-source fits, the more important question is how numerical error propagates through to the model hyper-parameters of our $R-L$ fit. In this section we give some recipes for estimating the scale of the numerical uncertainty relative to that of the statistical uncertainty, and demonstrate the statistical certainty dominates by two orders of magnitude.

First, note that the marginal likelihood of the model is:
\begin{equation}    
    \mathcal{L} \left( \left\{d^{n}\right\}\middle \vert  \Lambda , f\right)
    \propto
    \prod_{n} \left[ 
        f\cdot \BF{Lag}\left(d^{n} \vert \Lambda \right)+\left(1-f\right)
    \right]
    .
\end{equation}
Taking a simple gradient relative to an estimate of $\BF{Lag}\left(d^{n} \vert \Lambda \right)$, we get:
\begin{eqnarray}
    \frac{\partial \mathcal{L}\left(\left\{d^{n}\right\}\middle \vert  \Lambda ,f\right)}{\partial \left(\BF{Lag}\left(d^{n} \vert \Lambda \right)\right)}
    &\propto &
    \mathcal{L}\left(\left\{d^{n}\right\}\middle \vert  \Lambda ,f\right)\times \frac{f}{f\cdot \BF{Lag}\left(d^{n} \vert \Lambda \right)+\left(1-f\right)}
    \\&=& \nonumber
    \ \mathcal{L}\left(\left\{d^{n}\right\}\middle \vert  \Lambda ,f\right)\frac{P_\mathrm{Lag}^{n}}{\BF{Lag}\left(d^{n} \vert \Lambda \right)}
\end{eqnarray}
And the sensitivity to the $R-L$-parameters is:
\begin{eqnarray}
    \frac{\partial \mathcal{L}\left(\left\{d^{n}\right\}\middle \vert  \Lambda ,f\right)}{\partial \Lambda}
    &\propto &
    \prod_{n} \left[
    f\cdot \BF{Lag}\left(d^{n} \vert \Lambda \right)+\left(1-f\right) \right]
    \\ &\;\;\times& \nonumber
    \sum_{m} \frac{f\frac{\partial }{\partial \Lambda }\BF{Lag}\left(d^{m} \vert \Lambda \right)}{f\cdot \BF{Lag}\left(d^{m} \vert \Lambda \right)+\left(1-f\right)}
    \\&=& \nonumber
    \mathcal{L}\left(\left\{d^{n}\right\}\middle \vert  \Lambda ,f\right)\sum_{m} \frac{f\frac{\partial }{\partial \Lambda }\BF{Lag}\left(d^{m} \vert \Lambda \right)}{f\cdot \BF{Lag}\left(d^{m} \vert \Lambda \right)+\left(1-f\right)}
    \\&=& \nonumber
    \mathcal{L}\left(\left\{d^{n}\right\}\middle \vert  \Lambda ,f\right)\sum_{m} P_\mathrm{Lag}^{m}\left(\Lambda \right) \frac{\frac{\partial }{\partial \Lambda}\BF{Lag}\left(d^{m} \vert \Lambda \right)}{\BF{Lag}\left(d^{m} \vert \Lambda \right)} 
    \\&\approx& \nonumber
    \mathcal{L}\left(\left\{d^{n}\right\}\middle \vert  \Lambda ,f\right)\sum_{m}  \frac{P_\mathrm{Lag}^{m}\left(\Lambda \right)}{{\Delta }_{\mathrm{stat}}\Lambda }\frac{{\Delta }_{\Lambda }\BF{Lag}\left(d^{m} \vert \Lambda \right)}{\BF{Lag}\left(d^{m} \vert \Lambda \right)}\ 
    ,
\end{eqnarray}
where in the last line we have made the finite difference approximation that:
\begin{equation}
    \frac{\partial }{\partial \Lambda }\BF{Lag}\left(d^{m} \vert \Lambda \right)\approx \frac{{\Delta }_{\Lambda }\BF{Lag}\left(d^{m} \vert \Lambda \right)}{{\Delta }_{\mathrm{stat}}\Lambda }
\end{equation}
Where ${\Delta }_{\Lambda }\BF{Lag}\left(d^{m} \vert \Lambda \right)$ is the error in source $m$’s Bayes factor due to statistical uncertainty in the our fitted R-L parameters, ${\Delta }_{\mathrm{stat}}\Lambda$.

We are concerned with the possible change in the R-L parameters due to a numerical change in the Bayes factor, i.e. the size of ${\Delta }_{BF}\Lambda$  with respect to some error ${\Delta }_{\mathrm{num}}\BF{Lag}\left(d^{n} \vert \Lambda \right)$. This is approximated by a derivative:
\begin{equation}
    \frac{\partial \left(\Lambda \right)}{\partial \left(\BF{Lag}\left(d^{n} \vert \Lambda \right)\right)}\approx \frac{{\Delta }_{BF}\Lambda }{{\Delta }_{\mathrm{num}}\BF{Lag}\left(d^{n} \vert \Lambda \right)}
    ,
\end{equation}
which is approximated by the ratio of the partial derivatives we took above:
\begin{equation}
    \frac{{\Delta }_{BF,n}\Lambda }{{\Delta }_{\mathrm{num}}\BF{Lag}\left(d^{n} \vert \Lambda \right)}=\frac{\frac{P_\mathrm{Lag}^{n}}{\BF{Lag}\left(d^{n} \vert \Lambda \right)}}{\sum_{m} P_\mathrm{Lag}^{m}\left(\Lambda \right)\times \frac{1}{{\Delta }_{\mathrm{stat}}\Lambda }\frac{{\Delta }_{\Lambda }\BF{Lag}\left(d^{m} \vert \Lambda \right)}{\BF{Lag}\left(d^{m} \vert \Lambda \right)}}
    .
\end{equation}

With some rearranging we arrive at an expression for the relative impact of statistical and numerical uncertainty in our R-L parameter derivations:
\begin{equation}
    \frac{{\Delta }_{BF,n}\Lambda }{{\Delta }_{\mathrm{stat}}\Lambda }=P_\mathrm{Lag}^{n}{\left(\sum_{m} P_\mathrm{Lag}^{m}\left(\Lambda \right)\times \frac{{\Delta }_{\Lambda }\BF{Lag}\left(d^{m} \vert \Lambda \right)}{\BF{Lag}\left(d^{m} \vert \Lambda \right)}\right)}^{-1}\frac{{\Delta }_{\mathrm{num}}\BF{Lag}\left(d^{n} \vert \Lambda \right)}{\BF{Lag}\left(d^{n} \vert \Lambda \right)}\ 
    .
\end{equation}
This impact goes to zero if a source has either high numerical precision in in Bayes factor integral or strongly does not contain a lag. This is intuitive: sources not associated with the R-L relation do not impact its parameters.
We can estimate the uncertainties by:
\begin{eqnarray}
    \frac{{\Delta }_{\mathrm{num}}\BF{Lag}\left(d^{n} \vert \Lambda \right)}{\BF{Lag}\left(d^{n} \vert \Lambda \right)}
    &\approx& \frac{1}{\sqrt{ESS_{2}}},
    \\
    \frac{{\Delta }_{\Lambda }\BF{Lag}\left(d^{m} \vert \Lambda \right)}{\BF{Lag}\left(d^{m} \vert \Lambda \right)}
    &\approx&
    \frac{\BF{Lag}\left(d^{m} \vert {\Lambda }^{\mathrm{hi}}\right)-\BF{Lag}\left(d^{m} \vert {\Lambda }^{\mathrm{lo}}\right)}{\BF{Lag}\left(d^{m} \vert \widehat{\Lambda }\right)}
\end{eqnarray}

We find that, for all sources, the approximate numerical error is small, with the largest being $\tilde0.3\%$ of the statistical error. Taking the square sum error, i.e.$ \sqrt{\sum_{n} {\left(\frac{{\Delta }_{BF,n}\Lambda }{{\Delta }_{\mathrm{stat}}\Lambda }\right)}^{2}}$, we estimate a total numerical uncertainty of $<1\%$ for all lines. For future-proofing reasons, we note that, as ${\Delta }_{\Lambda }\BF{Lag}\left(d^{m} \vert \Lambda \right)\rightarrow 0$, i.e. as our $R-L$ relations become more precise, this uncertainty will begin to dominate. As a rough estimate of scaling, we expect the statistical uncertainty to scale again $\sqrt{N_\mathrm{Sources}}$ and the numerical uncertainty to scale with it, meaning the numerical uncertainty will begin to become significant (i.e. $\ge10\%$ of the statistical uncertainty) when our sample size increases by $\approx3-4\times$. As discussed in Section~\ref{app: num_futurescalability_popwide}, this should be a solvable problem by using $R-L$ informed priors as our pre-fit fiducial prior instead of the uniform priors we use here.

\subsection{Justifying the $\tmax$ Window Size}
\label{app: num_stability_tmax}
Our Monte Carlo integration scheme for calculating the $R-L$ parameter likelihoods in Section~\ref{sec: methods_practical_calc} reprocesses fiducial per-source fits using an uninformative prior, and in this work this prior is chosen to be uniform up to an upper bound of $\Delta t=\tmax$. By contrast, the $R-L$ informed log-normal prior extends to $\Delta t \rightarrow \infty$. To cover the evidence integral beyond the upper uniform prior bound, we make the assumption that $\mathcal{L}_\mathrm{Coup}(d^n\vert\Delta t^n)\approx\mathcal{L}_\mathrm{UnCoup}(d^n)$ as $\Delta t\rightarrow \infty$, as we discuss in Section~\ref{sec: methods_practical_calc}. Here we justify this approximation and show that our $\tmax$ upper bound is sufficient in most cases, as well as suggest some future improvements for sources with long timescales of variability.

\begin{figure}
    \centering
    \includegraphics[width=1.0\linewidth]{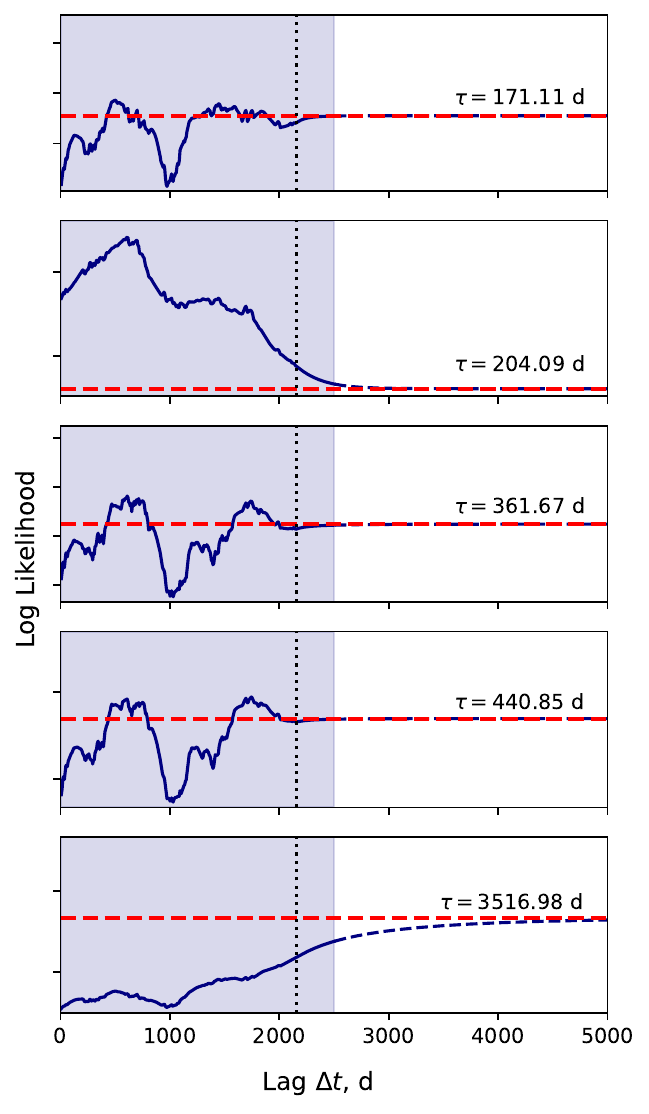}
    \caption{A demonstration of the lag log-likelihood curves for \ozdes \civ source $2970771568$ sources at posterior samples with all non-lag parameters drawn from random posterior values from the fiducial pre-fit. Vertical dashed lines mark the $\approx2160 \dayu$ survey length of \ozdes, the shaded blue region extends to the $\tmax$ upper bound of our uniform prior, and a horizontal orange dashed line shows the uncoupled likelihood. We can see that, as $\Delta t\rightarrow \infty$, the coupled likelihood converges to the uncoupled likelihood. \ozdes source, and that this convergence is almost entirely complete within the uniform prior except when the correlation timescale, $\tau$, is very long.}
    \label{fig: tmax_justify}
\end{figure}

The degeneracy between long lags and completely decoupled light curves holds when the margin between $\Delta t_\mathrm{Max}$ and the survey length is large compared to the coupling timescale, i.e. $(\Delta t_\mathrm{Max} - \mathrm{Survey \ Length})  / (\tau+b/2) \rightarrow  \infty$ (demonstrated in Figure~\ref{fig: tmax_justify}). This holds for many sources, but some of our light curve models yield long timescales which makes this a less accurate approximation. This should only have a major impact on sources that already have a lot of prior / posterior probability at $\Delta t> \tmax$ and are associated with the $R-L$ foreground distribution, i.e. have large $\BF{Lag}$. In our sample for fiducial fits, we find that there are no sources that meet both of these condition.

Many of our models suggest longer $\tau$ scales than the simple DRW model, and future surveys may examine more distant AGN in which the necessary $\Delta t_\mathrm{Max}$ is large. It is not practical to fit both the smooth unimodal flat region of the \qm{intermediate lag} region of $\mathrm{Survey \ Length} < \Delta t  < \Delta t_\mathrm{Max}$ and the complicated multimodal \qm{short lag} region $\Delta t < \mathrm{Survey \ Length}$ region simultaneously with nested sampling: the number of live points needed to fit that many modes is wasteful when the \qm{long lag} region is smooth and unimodal. If this becomes an issue, we suggest future works divide the posterior into a high resolution run (large $N_\mathrm{Live}$) for the short lag region, and a low resolution (small $N_\mathrm{Live}$) for the intermediate lag region, and then patch these NS chains together with weights proportional to the marginal likelihood of each region before Monte Carlo sampling. Alternately, non-sampling fitters like \litmus's \hescan or \sviscan scale better with prior width, but they are poorly suited to light curve models that are not differentiable.

\subsection{Scalability When Applying to Other Surveys}
\label{app: num_futurescalability}

In this work we apply our method to only the \ozdes sample, but as we discuss in Section~\ref{sec: widerimpact_future} this hierarchical modelling framework is deployable to any industrial-scale RM survey. There are two fronts on which we should consider the scalability of our method: how the pre-fitting of sources with uninformative priors can scale on a per-source basis, and how our hierarchical $R-L$ fit scales on a population-wide basis. We examine the scalability of the source-by-source fiducial pre-fitting in Appendix~\ref{app: num_futurescalability_persource}, and how the hierarchical population fitting will scale to larger samples and more complex models in Appendix~\ref{app: num_futurescalability_popwide}.

\paragraph{Per Source}
\label{app: num_futurescalability_persource}
To use this hierarchical method, we require fiducial source-by-source fits under an uninformative lag prior, which we acquire with nested sampling under uniform lag priors in this work. In this work, we find that the slower fits in the \ozdes sample take $\approx 25$ CPU hours to converge to good NS tolerance with $N_{Live}=2-3\times10^3$, with most fits converging in $1-2$ CPU hours. With each year of observation, we allow room for an additional seasonal aliasing peak, and thus an additional posterior mode. For proper exploration of the posterior space, the number of live points should scale linearly with the number of modes \citep{Albert_2020_JAXNS}. The number of evaluations (i.e. total numerical cost) of a nested sampling run scales linearly with the live points, meaning that this method, as applied to very long surveys, will have a per-source cost that similarly goes linearly with the number of years the survey runs for. It should be noted that the difficulty in fitting the lag posterior scales very poorly as we move into the high SNR regime, with the limit of very high precision data giving a number of effective modes that scales with the number of light curve epochs instead of the number of seasons \citep[see \papertwo and][]{Kroupa_2026b_likelihood}. 

In \papertwo we presented alternatives to nested sampling in the form of the \hescan and \sviscan, methods that scaled significantly better with the number of modes. To apply our Bayesian fitting in this work or in \papertwo and \paperthree to high SNR data, for example the type seen in accretion disk RM, we suggest avoiding nested sampling in favour of one of these accelerated methods. We note that the \hescan and \sviscan rely assume our likelihood functions are \sqm{well behaved} under differentiation. In this work, we find that the piece-wise GP kernel and transfer functions in Equations~\ref{eq: rc_covar}~and\ref{eq: rr_covar} (i.e. those corresponding to any tophat smoothing) run markedly slower and require more bespoke tuning of fitting parameters than the idealised cases demonstrated in \papertwo. If attempting to apply these fast methods to high SNR data, we suggest either using a continuous (not piecewise) light curve model, or else finding some other multi-modal friendly fitter that offers good support across the prior domain for the purposes of Monte Carlo integration. We make some suggestions for this in Appendix~\ref{app: num_stability_tmax}.

More expressive light curve models should not have a marked impact on the scalability of this method, provided the posterior distributions for any new parameters are unimodal. We note that the complexities of fitting the lag axis described in \papertwo become exponentially more troublesome as we try to fit multiple lags at once, and so using this method for simultaneous multi-band fitting may see a numerical bottleneck if using NS for the source-by-source pre fitting. A grid of \hescan test lags over multiple axes will allow us to sidestep this issue, but at the cost of needing to assume Gaussianity over non-lag axes.

\paragraph{Population Wide / Hierarchical Basis}
\label{app: num_futurescalability_popwide}
Our method for fitting hierarchical methods responds well to parallelisation; the pre-fitting of individual sources can be easily scaled, and the summation and product loops in the likelihood function of Equation~\ref{eq: marglik_finitesum} makes it embarrassingly parallel. In terms of numerical cost, we find that we can perform our fits for the mid $100$'s of \ozdes sources in less than a day with less than a dozen CPU cores when using the sampling conditions we outline in Section~\ref{sec: methods_practical_calc}. As we outline in Appendix~\ref{app: num_stability_RLpropagation}, our current fitting conditions will also remain numerically stable for an $\approx10-100\times$ larger AGN sample size. As such, our framework here should be immediately deployable to surveys like \sdss, \desi and \tides. 

As our population models become more complex or as sample sizes become significantly larger, we may find that the numerical uncertainty described in Appendix~\ref{app: num_stability_RLpropagation} begins to become significant as the ESS drops for load-bearing sources. In Appendix~\ref{app: num_futurescalability_persource} we suggest a few alternatives to NS to avoid this, but as a stop-gap measure we can continue to use NS in one of two ways. Firstly, we can simply increase the number of live points, though this cost scales linearly with desires ESS. Secondly, we can replace the fiducial uniform priors with a population informed prior, e.g. with a fiducial $R-L$ prior. Because this prior is cancelled out in the importance ratio, it leads to no confirmation bias in the final fit. We suggest this should only be done when we have some credible information about what the final population model will look like.

If the finite-sum marginal likelihood is numerically stable, this method should scale very well into more complex, high dimensional population models. Provided the number of hyper-parameters is in the low $10$'s, we can continue to use traditional MCMC methods. We note that our numerical formulation is differentiable, meaning high dimensional methods like HMC are viable for extremely complex models.

\section{Modelling Details \& Limitations}
\label{app: modelling}
In the main body of this work we present the most important points of our model procedure and its results, by and large using our fiducial model of a DRW continuum and a tophat smoothed response. In this appendix, we provide some details on how we gauge the sensitivity and reliability of these assumptions. In Appendix~\ref{app: modelling_modelsensitivity} we provide some more specific details on how sensitive our results are to our light curve model choice and what future improvements may be necessary in our description of the AGN variability. In Appendix~\ref{app: modelling_stationary} we examine the impact of light curves that cannot be described as a DRW, those that appear to be stationary or consistent with white noise, on our results.

\subsection{Sensitivity to Light Curve Model Choice \& Impact of Model Misspecification}
\label{app: modelling_modelsensitivity}

As with \paperthree, we can use the range of light curve models outlined in Section~\ref{sec: RM_stats_LCmodel} as a probe of our sensitive our results are to our model assumptions. We discuss this briefly in Section~\ref{sec: results_RL_constraints}, but offer more detail here and also examine what gap still exists between our models and the true AGN variability, and what influence this might have on the reverberation fractions measured in Section~\ref{sec: results_reverb}.

In our analysis, we find that the \hbeta sample gives consistent results for all model types. We suggest that this is owed to two things: firstly that the \hbeta response is \qm{well behaved}, i.e. with no complications of decoupling or non-reverberating variability, and secondly that \hbeta is associated with short and easy to read lags such that model mismatch is not as much of an important issue. By contrast, both \mgii and \civ need some level of smoothing to fit an $R-L$ relation, though for \mgii this can come by way of allowing for jitter in the continuum, which we noted in \paperthree smooths out the continuum's fit. The simple DRW fails for both of these lines because, while it can identify good fits, it cannot reject bad ones. This causes $f$ to have only a lower bound, accepting too many spurious recoveries and distorting the relation. For \civ, where lags are hard to identify at all, the jitter model is simply too vague to properly constrain anything. We show the constraints for all lines and all models in Table~\ref{tab: R-L_Constraints_allmodels} 

Our constraints on the reverberating fractions in Section~\ref{sec: results_reverb} have an important caveat: any misspecification of our light curve model will under-estimate $f$ by biasing our coupled model evidences downwards. In particular, if our model fails to describe the signal or noise processes, there will be apparent tensions between the light curves when we try to align them. This is of particular interest in examining the apparent lack of \mgii recoveries, as \mgii has well known difficulties in light curve calibration \& construction due to the presence of \feii contamination \citep[See ][for \ozdes's handling of this issue]{OzDES-Yu_2021, OzDES-Yu_2023}.

One way we can gauge how severely mismatch of our models may be affecting our estimates of the rate of coupled sources is to look at how the rate changes between different light curve models. In \paperthree found that $10-25\%$ of the \mgii sources demonstrated or preferred some noise in excess of their calibration error. Nonetheless, the jitter light curve model, which would absorb under-calibration of measurement uncertainties, gives constraints consistent with our fiducial model. This indicates that the paucity of $\mgii$ lags cannot be easily explained by under-estimating the calibration uncertainties in \ozdes's spectral reduction pipelines.

\begin{figure*}
    \centering
    \includegraphics[width=0.512\linewidth]{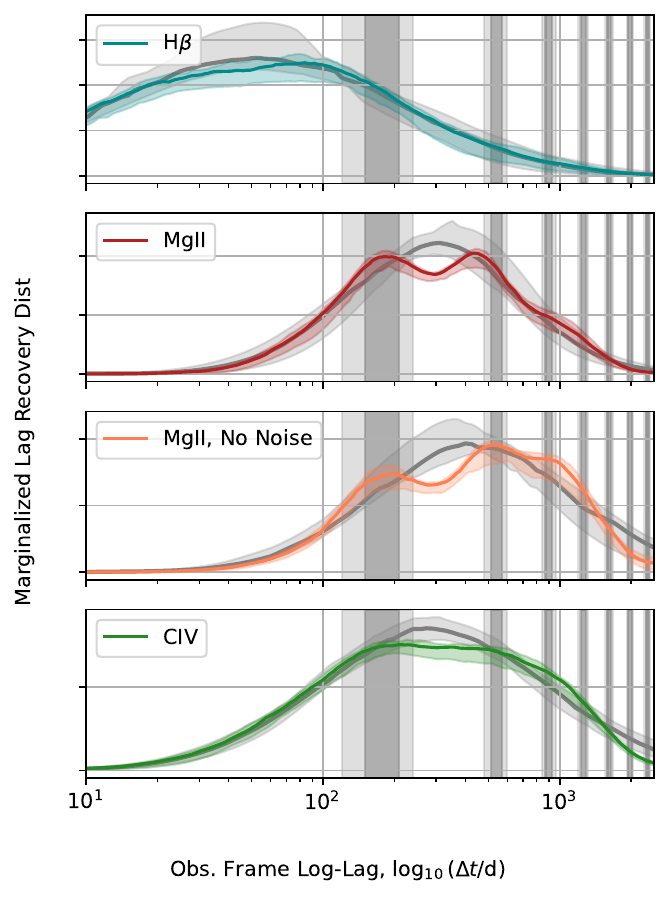}
    \includegraphics[width=0.468\linewidth, trim={1cm, 0, 0, 0}, clip]{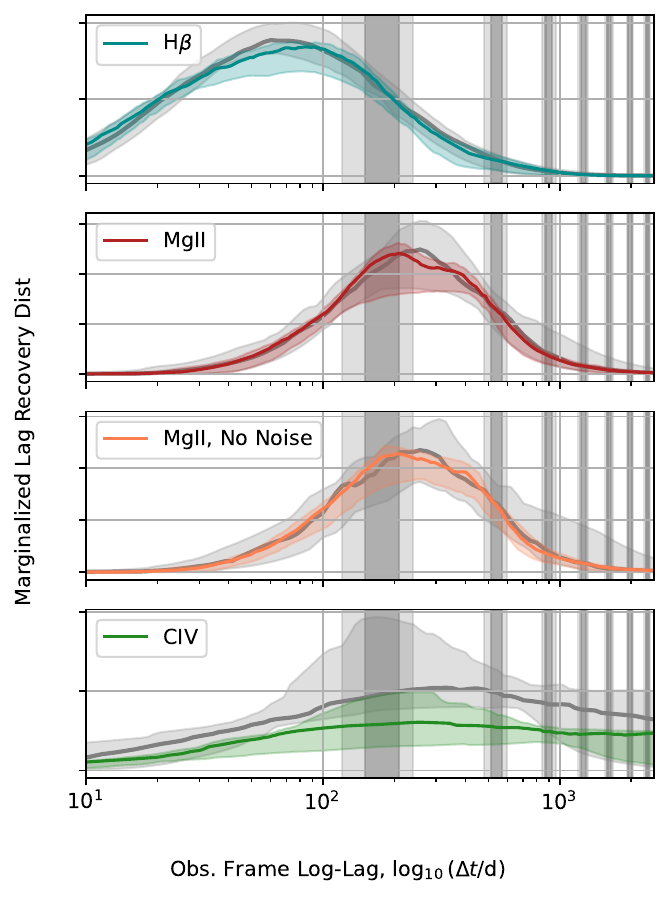}
    \caption{A comparison of the distribution for our recovered lags (coloured) to the distribution we expect from our population fits (grey, See Section~\ref{app: context_forecasting}). Both are plotted in the rest-frame on a log-scale and marginalised over the uncertainty in all population and source parameters. The left panel shows results for our fiducial tophat log smoothing model, and the right for the jitter model. The \qm{\mgii, No Noise} entry (row 3) are the results when marginalising out sources with stationary response light curves with the model in Section~\ref{app: modelling_stationary}. We can see that the \mgii distribution (middle two panels) are slightly suppressed near $\Delta t=360 \dayu$, causing them to over-represent the aliasing peaks either side (i.e. at $180\dayu and 540 \dayu$), which indicates some degree of model misspecification. \civ has the same effect to a lesser degree. Solid lines represent our median predicted distribution, while the shaded regions indicate the \pone and \ptwo percentile statistical uncertainties.}
    \label{fig: Compare_dists}
\end{figure*}

We can also use the fact that tensions between the light curves, and presumably the influence of model misspecification, is most apparent at lags where light curves overlap the most (i.e. on-season lags like $\Delta t\approx360 \dayu, 720 \dayu...$ where we can more easily compare the continuum and response). We can then gauge how badly our models are mismatched with our light curves by looking at how our lag recoveries \sqm{bunch up} at the aliasing peaks. Figure~\ref{fig: Compare_dists} shows this: our recoveries are under-represented compared to the prior in on-season lags for our fiducial model, which suggests at least some degree of light curve model mismatch is at play here. However, the jitter model (right columns in Figure~\ref{fig: Compare_dists}) does not show significant concentration at the aliasing peaks. Given the jitter model returns a similarly low $f_{\mgii}$, this tells us that the light curve model, though important, is not what is giving us such a low rate of recoveries.

Our work here demonstrates that (i) some amount of smoothing is needed in the transfer function for \mgii and \civ and (ii) we need better models of the non-reverberating component (i.e. the noise) for all lines. Noise characterisation of the response is of particular interest for \mgii, where we find many stationary light curves and know that the \feii contamination is likely to be impacting at least some of our light curve construction. A good first step in investigating these would be to fit to \sdss data with our method, which has a different \mgii pipeline but a similar physical footprint.

Note that in \paperthree, we found that the \des continuum light curves strongly preferred to have a short-timescale variance component of $\sim5-25\%$, corresponding to $2-30\dayu$, even when fit in isolation from the response light curves. The jitter model represents a worst case modelling of the departures from a simple DRW model, and, as we discuss in \paperthree, there may be significant gains to be made in GP-based RM by investigating more expressive models for the noise in the continuum in particular (e.g. outlier models to represent glitches / errors in data reduction or astrophysical transients like microlensing \citep{Best_2025_microlensing}). In that work we also suggested some ways to investigate the nature of this noise. 

\begin{table*}
    \centering
    \begin{tabular}{c|c|c|c|c|c|c}
        Line &  \makecell{Slope\\$\alpha$} & \makecell{Offset\\$\beta$} & \makecell{$\alpha$/$\beta$\\Corr.} & \makecell{Scatter\\$\sigma$} &  \makecell{Reverb.\\Frac. $f$} & \makecell{$\sigma/f$\\Corr.} \\\hline
        \multicolumn{7}{c}{Simple DRW}\\ \hline
        \hbeta & $0.51_{-0.15}^{+0.16}$ & $-0.40$ & $1.62_{-0.08}^{+0.08}$ & $0.38_{-0.06}^{+0.08}$ & $0.88_{-0.07}^{+0.06}$ & $0.33$\\\hline
        \mgii & $-0.06_{-0.69}^{+0.96}$ & $-0.44$ & $4.41_{-1.16}^{+0.82}$ & $1.22_{-0.39}^{+0.44}$ & $0.33_{-0.23}^{+0.35}$ & $0.13$\\\hline
        \civ & $0.25_{-0.24}^{+0.25}$ & $-0.79$ & $2.26_{-0.26}^{+0.26}$ & $0.74_{-0.11}^{+0.15}$ & $0.85_{-0.13}^{+0.09}$ & $0.34$\\\hline

        \multicolumn{7}{c}{Simple DRW + Jitter}\\ \hline
        \hbeta & $0.54_{-0.15}^{+0.16}$ & $-0.44$ & $1.55_{-0.08}^{+0.07}$ & $0.33_{-0.06}^{+0.07}$ & $0.84_{-0.07}^{+0.06}$ & $0.14$\\\hline
        \mgii & $0.36_{-0.44}^{+0.46}$ & $-0.65$ & $1.91_{-0.19}^{+0.17}$ & $0.25_{-0.08}^{+0.13}$ & $0.08_{-0.02}^{+0.03}$ & $0.17$\\\hline
        \civ & $0.46_{-0.72}^{+0.73}$ & $-0.28$ & $1.99_{-0.49}^{+0.62}$ & $0.94_{-0.56}^{+0.64}$ & $0.44_{-0.19}^{+0.30}$ & $0.76$\\\hline

    \end{tabular}
    \caption{Summaries of our $R-L$ parameter constraints for all lines when applying our Hierarchical model to the \ozdes sample as listed for the simple DRW and DRW + Jitter models described in Section~\ref{sec: RM_stats_LCmodel}, supplementary to the parameters for our fiducial model in Table~\ref{tab: R-L_Constraints}. Correlations assume Gaussian distributions for the parameters, and uncertainties are listed from the \pone and \ptwo percentiles relative to the posterior median. All fits are for the full sample without accounting for stationary light curves. The \hbeta relation is fit with a luminosity offset of $10^{44} \ \mathrm{erg/s}$, while the \mgii and \civ relations are fit for offsets of $10^{45} \ \mathrm{erg/s}$.}
    \label{tab: R-L_Constraints_allmodels}
\end{table*}

\subsection{Impact of Stationary Light Curves on Reverberating Fraction}
\label{app: modelling_stationary}
In \paperthree, we found that a large fraction of the apparent decoupling in the \mgii sample was due to response light curves that exhibited poor structure, i.e. that were consistent with white noise. Removing these poor SNR \mgii sources, we found that our \mgii reverberating fraction estimate increased to $\approx28$. We can examine this line of inquiry more thoroughly with our hierarchical model by adding a third hypothesis model to our summation in Equation~\ref{eq: marg_lik_all} that delineates between well structured and unstructured light curves. We say that some fraction of sources, $g$, present response light curves with no structure, and that $f$ now represents the fraction of the remaining sources that reverberate. In this way, the probability of a randomly selected source being coupled is $f\times(1-g)$. Our marginal likelihood from Equation~\ref{eq: marg_lik_all} now becomes:
\begin{eqnarray}
    \label{eq: marg_like_garble}
    P(\Lambda,f,g \vert D) \
    \propto
    \prod_{n} 
    \left[
    \begin{array}{ll}
     (1-g) \cdot \big( f\cdot \BF{Lag}^n(\Lambda)
     \\
     + (1-f) + f \cdot k^n(\Lambda) \big) \cdot \BF{Struc,Resp}^n + g
     \end{array}
     \right]
     .
\end{eqnarray}
Here, $\BF{Struc,Resp}^n$ is the Bayes factor between the uncoupled GP and GP-Noise models for source $n$, already available from our fiducial fits. Note that this collapses to our main modelling in Section~\ref{sec: methods} in the case of $g=0$.

\begin{figure}
    \centering    
    \includegraphics[width=1.0\linewidth, trim={0.0cm 0.0cm 0.0cm 0.0cm}]{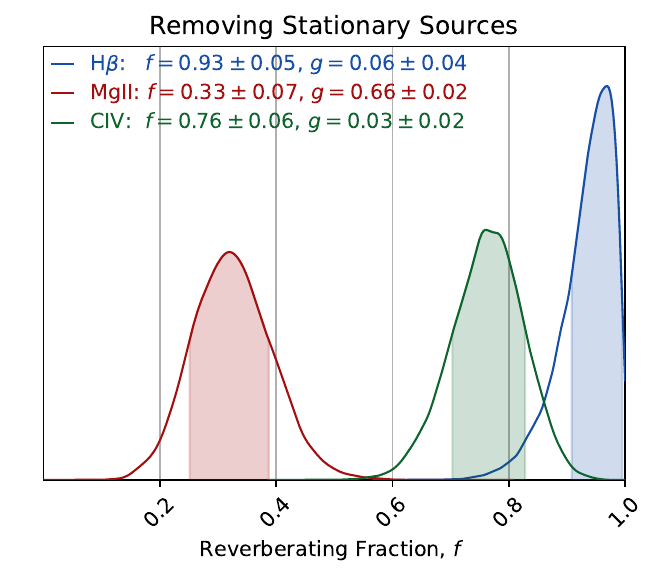}\\
    \caption{Posterior distributions for the underlying fraction of AGN, $f$, that exhibit lag-bearing coupled light curves between the continuum and response for all lines if we screen out stationary light curves as per our model in Equation~\ref{eq: marg_like_garble}. The legend also shows the constraints on $g$ for this model for each line.}
    \label{fig: reverb_frac_constraints_garbled}
\end{figure}

If we fit for $\Lambda$ with this model, we find $g_{\mgii}\approx66\%$ of light curves are stationary for \mgii, while , the reverberating fraction increases among the remaining high SNR sources to $f_{\mgii}=33\%\pm7\%$ to $0.33$ (see Figure~\ref{fig: reverb_frac_constraints_garbled}). In short, only $\approx1/3$ of our \mgii sample has well structured variability in their light curves, and of those only $\approx1/3$ confidently encode a lag. We discuss this in detail in Section~\ref{app: modelling_modelsensitivity}, but this conclusion is consistent over all of our well-fit light curve models. This is in line with our highest estimates in \paperthree, and suggests this is a fundamental limit on our lag recovery rates in \ozdes \mgii sample with its current data reduction pipeline. This does not separate out sources with no structure due to low variability, i.e. stationary constant light curves, and those that are highly variable, i.e. noisy white-noise like light curves. Similarly, it does not delineate between light curves that are catastrophically unstructured or those that demonstrate poor structure due to issues with the data.

Marginalising out noisy sources in this way has less impressive effects for \hbeta and \civ. Both see a slight increase in their reverberating fraction, but less than their statistical uncertainty, and have constraints that are consistent with $g=0$. Of note is that $f_{\hbetamath}$ is now consistent with $100\%$, indicating that the small number of poorly coupled sources are better explained by poor SNR than by a true confident detection of decoupling.

\section{This Work in the Broader RM Context}
\label{app: context}
In Section~\ref{sec: widerimpact} we discuss how this work should be contextualised in the broader field of reverberation mapping. In this appendix we provide more extensive statistical details on how our results compare to past industrial-scale surveys and $R-L$ fits, namely those of \sdss, past \ozdes works and the joint-analysis we made in \paperone. In Appendix~\ref{app: context_forecasting} we give some estimates of how next generation RM surveys like \tides or \desi and its succesors should expect their lag harvest rates to scale as they extend their temporal baseline.

\begin{figure}
    \centering
    \includegraphics[width=0.85\linewidth, trim={0.75cm 0.6cm 0.0cm 0.0cm}, clip]{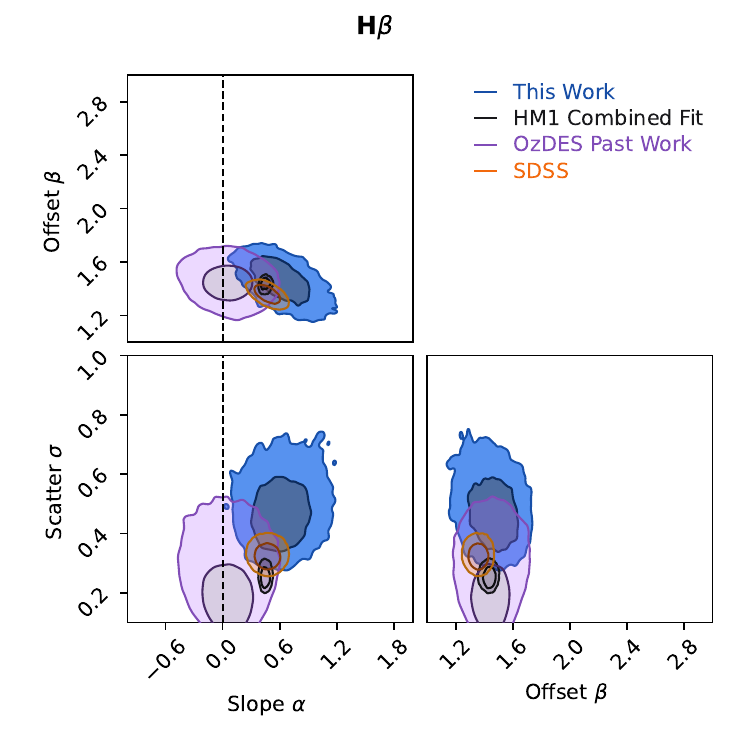}\\
    \includegraphics[width=0.85\linewidth, trim={0.75cm 0.6cm 0.0cm 0.0cm}, clip]{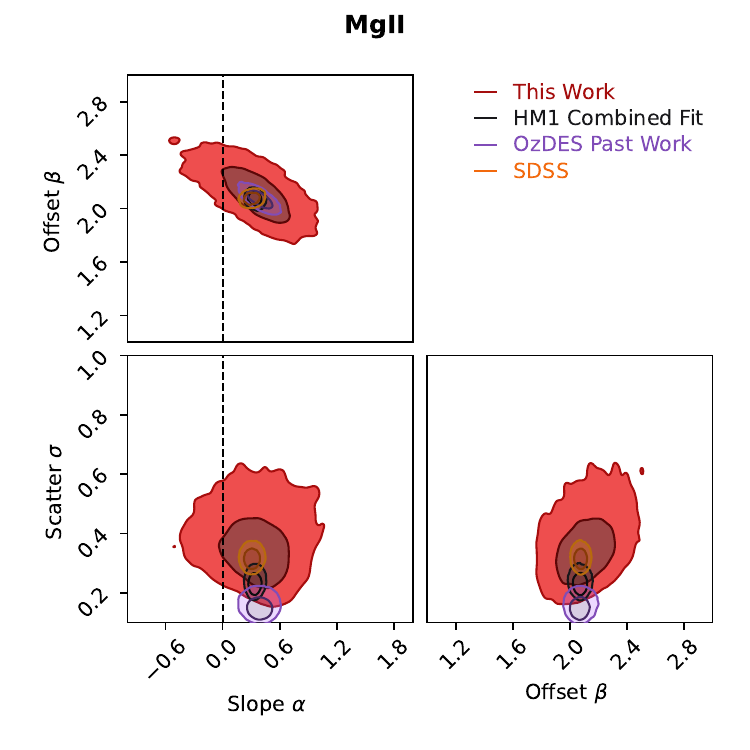}\\
    \includegraphics[width=0.85\linewidth, trim={0.75cm 0.6cm 0.0cm 0.0cm}, clip]{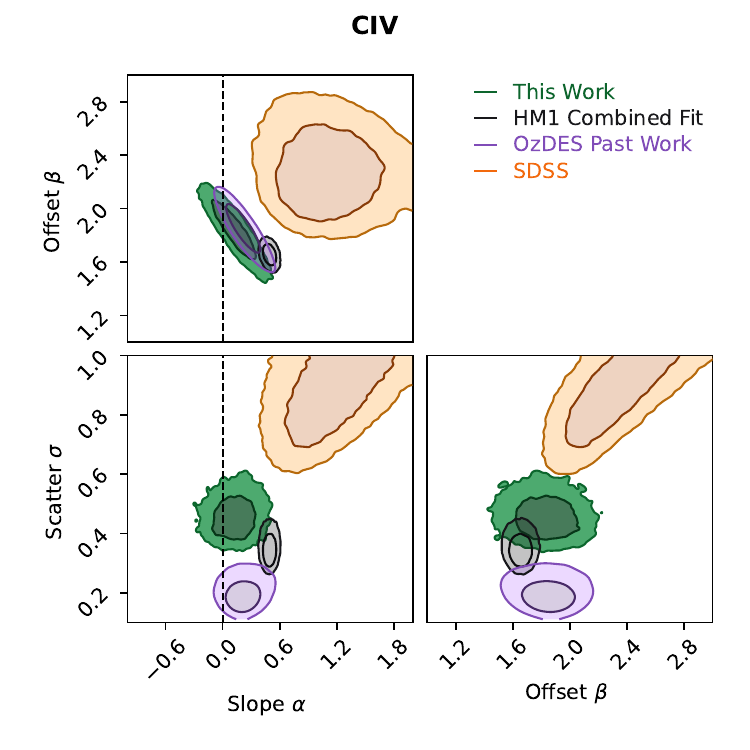}
    \label{fig: contours_compare}
    \caption{Posterior contours for our $R-L$ parameter constraints for all lines, compared to constraints using the cut-and-constrain approach using the results of \paperthree, the previous \javelin-based \ozdes lags, the \sdss data release of \citet{SDSS-Shen_2023} and the \ozdes-like subsample we took from that data in \paperone, as well as the final combined-survey $R-L$ relations we presented in \paperone.}
    \label{fig: contours_compare_lit}
\end{figure}

Figure~\ref{fig: contours_compare_lit} shows the $R-L$ constraints for our fiducial model choice as compared to $R-L$ constraints derived from other data sources. In \paperone we noted that the past lag samples of \ozdes and \sdss were in tension, most notably in their reported scatter in the $R-L$ plane, with \ozdes's past lag fits being markedly more tightly distributed. For the \hbeta relation, we find that our new $R-L$ results have a markedly higher scatter than those of \citet{OzDES-Malik_2023}, as well as a sharper slope and a higher offset. Though none of these changes are beyond statistical uncertainty, it brings the population fits of the \ozdes RM sample much more closely into alignment with the lags of \sdss as presented in \citet{SDSS-Shen_2023}.

\subsection{Forecasting Lag Recoveries for Future Surveys}
\label{app: context_forecasting}
\begin{figure}
    \centering
    \includegraphics[width=1.0\linewidth]{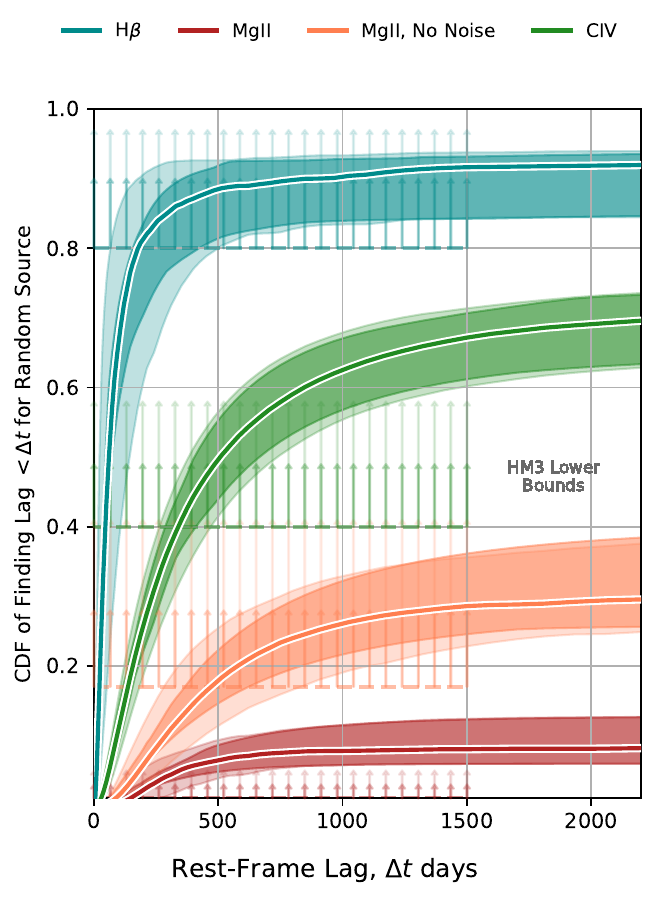}
    \caption{Our estimates of the probability of a randomly selected optical AGN from an \ozdes-like survey having a lag below a threshold $\Delta t_\mathrm{Ceil.}$, subdivided by line. The dark shaded regions are the \pone and \ptwo uncertainty bounds varying only the population parameters, the light shaded regions are the same but while also bootstrapping over the sample. The solid curves are the median probabilities after varying population properties and bootstrapping. The arrows below $1500 \dayu$ represent the conservative lower bounds for the reverberating fraction that we recovered in \paperthree. }
    \label{fig: completeness_estimates_plot}
\end{figure}

In addition to constraining the $R-L$ relation and our lag sample, our rest-frame lag distributions of our $R-L$ informed priors allows us to estimate the distribution of rest-frame lags in our sample. For other \ozdes-like high-redshift optical surveys, this allows us to put bounds on the maximum recovery rates for a survey of a particular length. We construct a completeness estimator by asking, for a randomly chosen source from $N$ sources, how likely that source is to have a lag below some rest-frame ceiling, $\Delta t_\mathrm{Ceil.}$:
\begin{equation}
    P(\mathrm{\Delta t<\Delta t_\mathrm{Ceil.}}\vert \Lambda,f)
    =
    \frac{f}{N_\mathrm{Sources}}
    \sum_n{
    \int_0^{\Delta t_\mathrm{Ceil.}}{\pi^n(\Delta t^n\vert \Lambda)d\Delta t^n}
    }
    ,
    \label{eq: plag_sum}
\end{equation}
where the summation here is over the luminosities and redshifts of the sources in the \ozdes sample for a particular line. To quantify the statistical uncertainty in Equation~\ref{eq: plag_sum}, we vary the $R-L$ parameters, $\Lambda$, and reverberation fraction, $f$, within their posterior bounds. To account for the limited \ozdes sample size, we also bootstrap (re-sample with replacement) the sources in the summation over $n$. We show these completeness estimates in Figure~\ref{fig: completeness_estimates_plot}.

A typical rule of thumb is that a survey baseline must be $2-3\times$ the length of the observer frame lag for it to be observable \citep{OzDES-Penton_2021}. Assuming the \ozdes sample is representative of the sort of AGN RM candidates we have for an industrial-scale RM survey, this gives the probability that a randomly chosen RM candidate AGN will have a lag that we could observe. Using these constraints, we can estimate what fraction of targets in future surveys like \tides and \desi will exhibit measurable lags as they increase their baseline over time. The \hbeta curve levels out ($>80\%$ of maximum yield) at $\approx 500 \dayu$, meaning a $1500 \dayu$ ($\approx 4$ year survey) will begin to saturate the maximum number of possible lags. The predictions for \mgii, whether we use the entire sample or limit our selection to only high SNR light curves (see Appendix~\ref{app: modelling_stationary}) only begins to plateau at $\approx 1000 \dayu$ and \civ flattens out at even longer limits of $\approx 1500 \dayu$. This means that our surveys will need to extend for $1-2$ decades for us to observe the longest lags and push beyond the lag ceiling we discuss in \paperone. Until we do, we can only extrapolate from short-lag sources as we have done in this work.

\newpage

\onecolumn
\section{Table of Lags}
\label{app: final_sample}
\vspace{25pt}
Here we present our final sample of $347$ lags. For brevity, this table consists of only the targets with high quality lags, as measured by having a marginal probability of containing a lag of more than $50\%$ per Equation~\ref{eq: plag}. A full un-truncated sample of all $797$ samples is available in the online data release (see Section~\ref{sec: data_release}). All results here are using our fiducial light curve model tophat smoothing with no jitter. Masses are estimated per Equation~\ref{eq: RM_mass} using the virial factor of $\log_{10}(f)=0.62 \pm 0.07 \pm 0.31$ from \citet{SDSS-Shen_2023} (the first and second uncertainties here being statistical error and population scatter). The velocity dispersions are as presented in the data release of \paperone. 

\vspace{25pt}

\begin{longtable}{c|c|c|c|c|c|c|c|c|c}
    \centering
\ozdes ID 
& Line 
& Redshift 
& \makecell{$\log_{10}\left(\frac{L_\lambda}{ \left[\text{erg/s}\right]}\right)$} 
& \makecell{Marginal Lag\\Constraint$\Delta t \dayu$} 
& $\log_{10}(\BF{Lag})$ 
& $P_\mathrm{Lag}$
& \makecell{Rest-Frame\\Lag $\Delta t_\mathrm{Rest} \dayu$}
& \makecell{Vel. Disp.\\$\sigma_v$ (km/s)}
& \makecell{$\log_{10}$ of Est. \\ Mass($M_\odot$)} \\\hline

2925344542 & \hbeta & $0.37$ & $44.29$ & ${70.7}^{-26.5}_{+39.9}$ & $0.68$ & $98.4 \%$ & ${51.5}_{-19.3}^{+29.0}$ & $1622 \pm 9$ & ${7.74}_{-0.39}^{+0.39}$ \\\hline
2970738404 & \hbeta & $0.45$ & $44.04$ & ${138.0}^{-41.3}_{+72.4}$ & $0.87$ & $98.9 \%$ & ${95.0}_{-28.4}^{+49.8}$ & $1712 \pm 7$ & ${8.07}_{-0.38}^{+0.36}$ \\\hline
2925598080 & \hbeta & $0.63$ & $44.46$ & ${35.9}^{-15.2}_{+48.1}$ & $0.25$ & $96.2 \%$ & ${22.1}_{-9.3}^{+29.5}$ & $1748 \pm 13$ & ${7.46}_{-0.44}^{+0.45}$ \\\hline
2940443426 & \hbeta & $0.63$ & $44.38$ & ${59.1}^{-39.0}_{+49.5}$ & $0.66$ & $98.2 \%$ & ${36.2}_{-23.9}^{+30.3}$ & $1613 \pm 17$ & ${7.55}_{-0.53}^{+0.44}$ \\\hline
2925592857 & \hbeta & $0.49$ & $44.16$ & ${92.2}^{-58.1}_{+134.8}$ & $0.78$ & $98.6 \%$ & ${62.0}_{-39.1}^{+90.6}$ & $1917 \pm 22$ & ${7.95}_{-0.56}^{+0.47}$ \\\hline
2970758215 & \hbeta & $0.41$ & $44.63$ & ${105.1}^{-71.3}_{+164.5}$ & $0.30$ & $93.9 \%$ & ${74.8}_{-50.7}^{+117.0}$ & $3096 \pm 55$ & ${8.42}_{-0.57}^{+0.56}$ \\\hline
2925577248 & \hbeta & $0.38$ & $43.69$ & ${15.4}^{-11.2}_{+72.0}$ & $0.55$ & $96.7 \%$ & ${11.1}_{-8.1}^{+52.1}$ & $1657 \pm 16$ & ${7.09}_{-0.60}^{+0.71}$ \\
\end{longtable}

\end{document}